\documentclass[12pt]{article}
\usepackage{geometry}                
\usepackage{graphicx}
\usepackage{amssymb}
\usepackage{amsmath}
\usepackage{epstopdf}

\begin{document}
\begin{titlepage}

\vskip 5em

\begin{center}
{\bf \LARGE
Criticality, Scaling and Chiral Symmetry Breaking in External Magnetic Field}
\vskip 4em

\centerline{\bf Veselin G. Filev}

\bigskip

  \centerline{\it Department of Physics and Astronomy, University of
Southern California}
\centerline{\it Los Angeles, CA 90089-0484, U.S.A.}

\centerline{\small \tt  filev@usc.edu}
\bigskip

\bigskip
\bigskip

\vskip 4em
\begin{abstract}
We consider a D7-brane probe of AdS$_{5}\times S^5$ in the presence of pure gauge $B$-field. The dual gauge theory is flavored Yang-Mills theory in external magnetic field. We explore the dependence of the fermionic condensate on the bare quark mass $m_{q}$ and study the discrete self-similar behavior of the theory near the origin of the parametric space. We calculate the critical exponents of the bare quark mass and the fermionic condensate. A study of the meson spectrum 
supports the expectation based on thermodynamic considerations that at zero bare quark mass the stable phase of the theory is a chiral symmetry breaking one. Our study reveals the self-similar structure of the spectrum near the critical phase of the theory, characterized by zero fermionic condensate and we calculate the corresponding critical exponent of the meson spectrum.

 \end{abstract}
\end{center}
\end{titlepage}

\section{Introduction}
The existence of  gauge/sting dualities has been anticipated from the very beginning of these fields, based on the striking similarity
    between the large N t'Hooft limit and the genus expansion in string theory. Further insights were revealed with the discovery of
     D-branes, their description in terms of DBI action and identification as sources of the known p-brane solutions in super gravity.
     However it was not until Maldacena conjectured his decoupling limit when significant progress was made and the AdS/CFT correspondence
     emerged \cite{Aharony:1999ti} as a manifestation of the gauge/string duality providing holographic description of super gravity on
     $AdS_{5}\times S^5$ space in terms of $\cal N$=4 SUSY Yang-Mills  theory living at the asymptotic boundary. The observed correspondence
     was conjectured to hold for the full string theory on any asymptotically $AdS_5\times S^5$ background. One of the most remarkable
     features of the correspondence is that it is a strong-weak correspondence and that it can give us tools to explore the strongly coupled
     regimes of the Yang-Mills theory.
     
    In  recent years progress has been made towards the study of  matter in fundamental representation in the context of AdS/CFT correspondence. One way
    to achieve this is by introducing space filling flavor D7-branes in the probe limit \cite{Karch:2002sh} and in order to keep the probe
     limit valid the condition $N_{f} \ll N_{c}$ is imposed. The fundamental strings stretched between the stack of $N_{c}$ D3 branes and
     the flavor $N_{f}$ D7-branes give rise to $\cal N$=2 hypermultiplet, the separation of the D3 and D7 branes in the transverse
     directions corresponds to the mass of the hypermultiplet, the classical shape of the D7-brane encodes the value of the fermionic
     condensate and its quantum fluctuations describe the light meson spectrum of the theory \cite{Kruczenski:2003be}. This technique for
      introducing fundamental matter has been widely employed in different backgrounds. Of particular interest was the study of non
      supersymmetric backgrounds and phenomena such as spontaneous chiral symmetry breaking. These phenomena were first studied in this context in \cite{Babington:2003vm}, where the authors developed an appropriate numerical technique. In  recent years this approach received further development, and has proven itself as powerful tool for the exploration of confining gauge theories, in particular, for the description of their thermodynamic properties or for the building of phenomenological models relevant to QCD\cite{Kruczenski:2003uq}-\cite{Myers:2007we}.

The paper is organized as follows:\\
In the second section we review the method of introducing magnetic field to the theory, employed in \cite{Filev:2007gb}. We describe the basic properties of the D7 brane embedding and the thermodynamic properties of the dual gauge theory, in particular the dependence of the fermionic condensate on the bare quark mass. We describe the spontaneous chiral symmetry breaking caused  by the external magnetic field and comment on the spiral structure in the condensate vs. bare quark mass diagram.

The third section contains our main results and splits into two parts:\\
The first part is dedicated to the detailed study of the spiral structure described in \cite{Filev:2007gb}. We perform analysis similar to the one considered in \cite{Frolov:2006tc} for the study of merger transitions and calculate the critical exponents of the bare quark mass and the fermionic condensate. We also describe the discrete self-similarity of the spiral and calculate the scaling factor characterizing it. 

In the second part of this section we consider the meson spectrum of the states corresponding to the spiral. First we study the critical embedding corresponding to the center of the spiral and reveal an infinite tower of tachyonic states organized in a decreasing geometrical series. Next we consider the dependence of the meson spectrum on the bare quark mass and confirm the expectations based on thermodynamic considerations that only the lowest branch of the spiral is stable. We observe that at each turn of the spiral there is one new tachyonic state. We comment on the self-similar structure of the spectrum and calculate the critical exponent of the meson mass. We also consider the spectrum corresponding to the lowest branch of the spiral and for a large bare quark mass reproduce the result for pure ${\cal N}=2$ Supersymmetric Yang Mills Theory obtained in \cite{Kruczenski:2003be}. 

We end with a short discussion of our results and the possible directions of future study.

\section{Fundamental matter in external magnetic field}
In this section we briefly review the method of introducing external magnetic field to the theory considered in \cite{Filev:2007gb}  and the basic properties of the D7-brane probe in this background. We also review the properties of the corresponding dual theory and the effect that the external magnetic field has on it. 
\subsection{Basic Configuration}

Let us consider the $AdS_{5} \times S^5$ geometry describing the near horizon geometry of a stack of $N_{c}$ extremal D3-branes.
\begin{eqnarray}
ds^2=\frac{u^2}{R^2}(-dx_{0}^2+d\vec x^2)+R^2\frac{du^2}{u^2}+R^2d\Omega_{5}^2,\label{AdS}\\
g_{s}C_{(4)}=\frac{u^4}{R^4}dx^0\wedge dx^1\wedge dx^2 \wedge dx^3,\nonumber\\\
e^\Phi=g_s,\nonumber\\
R^4=4\pi g_{s}N_{c}\alpha'^2\nonumber\ .
\end{eqnarray}

In order to introduce fundamental matter we first rewrite the metric in the following form :
\begin{eqnarray}
ds^2&=&\frac{\rho^2+L^2}{R^2}[ - dx_0^2 + dx_1^2 +dx_2^2 + dx_3^2 ]+\frac{R^2}{\rho^2+L^2}[d\rho^2+\rho^2d\Omega_{3}^2+dL^2+L^2d\phi^2],\nonumber\\
d\Omega_{3}^2&=&d\psi^2+\cos^2\psi d\beta^2+\sin^2\psi d\gamma^2, \label{geometry1}
\end{eqnarray}
where $\rho, \psi, \beta,\gamma$ and $L,\phi$ are polar coordinates in the transverse ${\cal R}^4$ and ${\cal R}^2$ respectively, satisfying: $u^2=\rho^2+L^2$.
Next we use $x_{0,1,2,3},\rho,\psi,\beta,\gamma$ to parametrise the world volume of the D7-brane and consider the following ansatz
\cite{Karch:2002sh} for it's embedding:
\begin{eqnarray}
\phi\equiv const,\\
L\equiv L(\rho)\ .\nonumber \label{ansatzEmb}
\end{eqnarray}
Leading to the following form of the induced metric:
\begin{equation}
d\tilde s=\frac{\rho^2+L(\rho)^2}{R^2}[ - dx_0^2 + dx_1^2 +dx_2^2
+dx_3^2]+\frac{R^2}{\rho^2+L(\rho)^2}[(1+L'(\rho)^2)d\rho^2+\rho^2d\Omega_{3}^2]\ .\label{inducedMetric}
\end{equation}
Now let us consider the NS part of the general DBI action:
\begin{eqnarray}
S_{DBI}=-\frac{\mu_{7}}{g_s}\int\limits_{{\cal M}_{8}}d^{8}\xi det^{1/2}(P[G_{ab}+B_{ab}]+2\pi\alpha' F_{ab})\ . \label{DBI}
\end{eqnarray}

Here $\mu_{7}=[(2\pi)^7)\alpha'^4]^{-1}$ is the D7-brane tension, $P[G_{ab}]$ and $P[B_{ab}]$ are the induced metric and $B$-field on the D7-brane's world volume, while $F_{ab}$ is its gauge field. A simple way to
introduce magnetic field would be to consider pure gauge $B$-field along the "flat" directions
 of the geometry $x_{0}-x_{3}$ corresponding to the D3-branes world volume:
\begin{equation}
B^{(2)}= Hdx_{2}\wedge dx_{3}\ . \label{ansatz}
\end{equation}
The constant $H$ is proportional to the magnetic component of the EM field. Note that since the $B$-field is a pure gauge $dB=0$ the corresponding background is still a solution to the supergravity equations. On the other side the gauge field $F_{ab}$ comes in next order in $\alpha'$ expansion compared to the metric and the $B$-field components. Therefore to study the classical embedding of the D-brane one can leave only the $G_{ab}+B_{ab}$ part of the DBI-action. It was argued in \cite{Filev:2007gb} that one can consistently satisfy the constraints imposed on the classical embedding resulting from integrating out $F_{ab}$. The resulting effective lagrangian is:
\begin{equation}
{\cal L}=-\frac{\mu_{7}}{g_s}\rho^3\sin\psi\cos\psi\sqrt{1+L'^2}\sqrt{1+\frac{R^4H^2}{(\rho^2+L^2)^2}}\ .
\label{lagrangian}
\end{equation}
The equation of motion for the profile $L_0(\rho)$ of the D7-brane is given by:
\begin{equation}
\partial_{\rho}\left(\rho^3\frac{L_0'}{\sqrt{1+L_0'^2}}\sqrt{1+\frac{R^4H^2}{(\rho^2+L_0^2)^2}}\right)+ \frac{\sqrt{1+L_0'^2}}{\sqrt{1+\frac{R^4h^2}{(\rho^2+L_0^2)^2}}}\frac{2\rho^3L_0R^4H^2}{(\rho^2+L_0^2)^3}=0\ .
\label{eqnMnL}
\end{equation}
As expected for large $(L_0^2+\rho^2) \to \infty$ or $H \to 0$, we get the equation for the pure AdS$_{5}\times S^5$ background \cite{Karch:2002sh}:
\begin{eqnarray*}
\partial_{\rho}\left(\rho^3\frac{L_0'}{\sqrt{1+L_0'^2}}\right)=0\ .
\end{eqnarray*}
Therefore the solutions to equation (\ref{eqnMnL}) have the following behavior at infinity:
\begin{equation}
L_0(\rho)=m+\frac{c}{\rho^2}+\dots,
\end{equation}
where the parameters $m$ (the asymptotic separation of the D7- and D3- branes) and $c$ (the degree of bending of the D7-brane) are related to the bare quark mass $m_{q}=m/2\pi\alpha'$ and the fermionic condensate $\langle\bar\psi\psi\rangle\propto -c$ respectively \cite{Kruczenski:2003uq}. We have provided derivation of these relations in Appendix A. As we shall see below, the presence of the external magnetic field and its effect on the dual SYM  provide a non vanishing value for the fermionic condensate, furthermore the theory exhibits chiral symmetry breaking.

   Now notice that $H$ enters in (\ref{lagrangian}) only through the combination $H^2R^4$. The other natural scale is the asymptotic separation $m$. It turns out that different physical configurations can be studied in terms of  the ratio $\tilde m^2={m^2}/{(H R^2)}$: Once the $\tilde m$ dependence of our solutions  are known, the $m$ and $H$ dependence follows. Indeed let us introduce dimensionless variables {\it via}:
  \begin{eqnarray}
  \rho=R\sqrt{H}\tilde\rho\ , \quad
  L_0=R\sqrt{H}\tilde L\ , \quad
  L_0'(\rho)=\tilde L'(\tilde\rho)\ .\label{cordchange}
  \end{eqnarray}
The equation of motion (\ref{eqnMnL})  then takes the form:
\begin{equation}
\partial_{\tilde\rho}\left(\tilde\rho^3\frac{\tilde L'}{\sqrt{1+{\tilde L}'^2}}\sqrt{1+\frac{1}{(\tilde\rho^2+\tilde L^2)^2}}\right)+ \frac{\sqrt{1+\tilde L'^2}}{\sqrt{1+\frac{1}{(\tilde\rho^2+\tilde L^2)^2}}}\frac{2\tilde\rho^3\tilde L}{(\tilde\rho^2+\tilde L^2)^3}=0\ .
\label{eqnMnLD}
\end{equation}
The solutions for $\tilde L(\tilde\rho)$ can be expanded again to:
\begin{equation}
\tilde L(\tilde\rho)=\tilde m+\frac{\tilde c}{\tilde\rho^2}+\dots, \label{ExpansionD}
\end{equation}
and using the transformation (\ref{cordchange}) we can get:
\begin{equation}
c=\tilde c R^3H^{3/2} \label{Hdepend}\ .
\end{equation}

\subsection{Properties of the Solution}
The properties of the solution have been explored in \cite{Filev:2007gb}, both numerically and analytically, when possible. Let us briefly review the main results.

For weak magnetic field $H$ and non-zero bare quark mass $m$ it was shown that the theory develops a fermionic condensate:
\begin{equation}
\langle\bar\psi\psi\rangle \propto -c =-\frac{R^4}{4m}H^2\ , \label{condSmA}
\end{equation}
or using dimensionless variables:
\begin{equation}
\tilde c=\frac{1}{4\tilde m} \label{1/m}\ .
\end{equation}

\begin{figure}[h] 
   \centering
   \includegraphics[width=10cm]{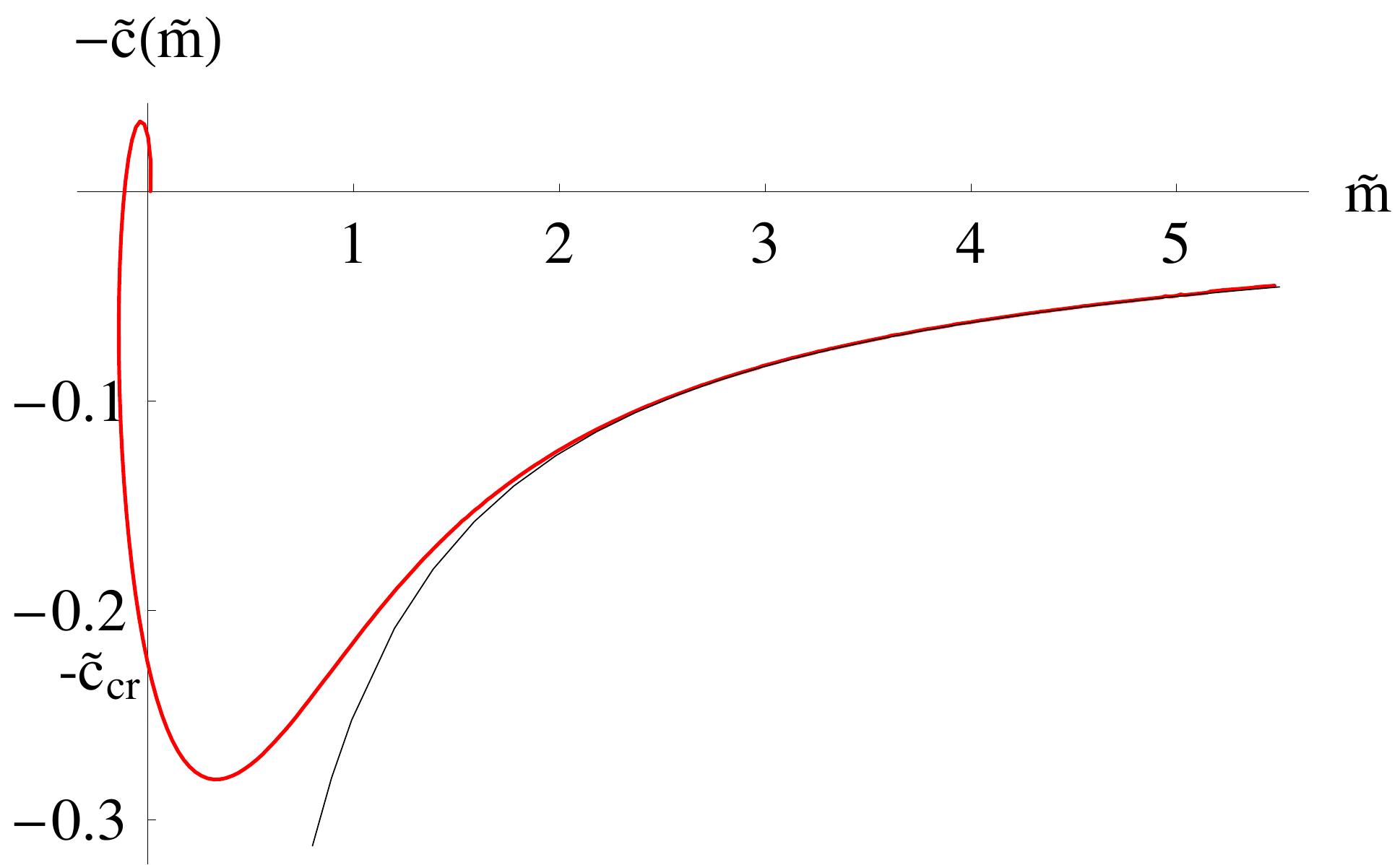}
   \caption{\small The black line corresponds to (\ref{1/m}), one can observe that the analytic result is valid for large $\tilde m$.
    It is also evident that for $\tilde m=0$ $\langle\bar\psi\psi\rangle\neq0$. The corresponding value of the condensate is $\tilde c_{\rm cr}=0.226$.}  \label{fig:fig1}
\end{figure}

The case of strong magnetic field $H$ can be explored by numerically solving equation (\ref{eqnMnLD}), it is convenient to use initial conditions in the IR as has been recently discussed in the literature \cite{Albash:2006ew}, \cite{Albash:2006bs}. We used the boundary condition  $\tilde L'(\tilde\rho)\vert_{\tilde\rho=0}=0$. We used  shooting techniques to generate the embedding of the D7 for a wide range of $\tilde m$. Having done so we expanded numerically the solutions for $\tilde L(\tilde\rho)$ as in equation (\ref{ExpansionD}) and generated the points in the  $(\tilde m,-\tilde c)$ plane corresponding to the solutions. The resulting plot is presented in figure~\ref{fig:fig1}.

As one can see there is a non zero fermionic condensate for zero bare quark mass and hence there is a Spontaneous Breaking of the  Chiral Symmetry. The corresponding value of the condensate is $\tilde c_{\rm cr}=0.226$. It is also evident that the analytical expression for the condensate (\ref{1/m}) that we got in the previous section is valid for large $\tilde m$, as expected. Now using equation (\ref{Hdepend}) we can deduce the dependence of $c_{\rm cr}$ on $H$:
\begin{equation}
c_{\rm cr}=\tilde c_{\rm cr}R^3H^{3/2}=0.226R^3H^{3/2}\ . \label{Ccr}
\end{equation}

\begin{figure}[h] 
   \centering
   \includegraphics[width=9cm]{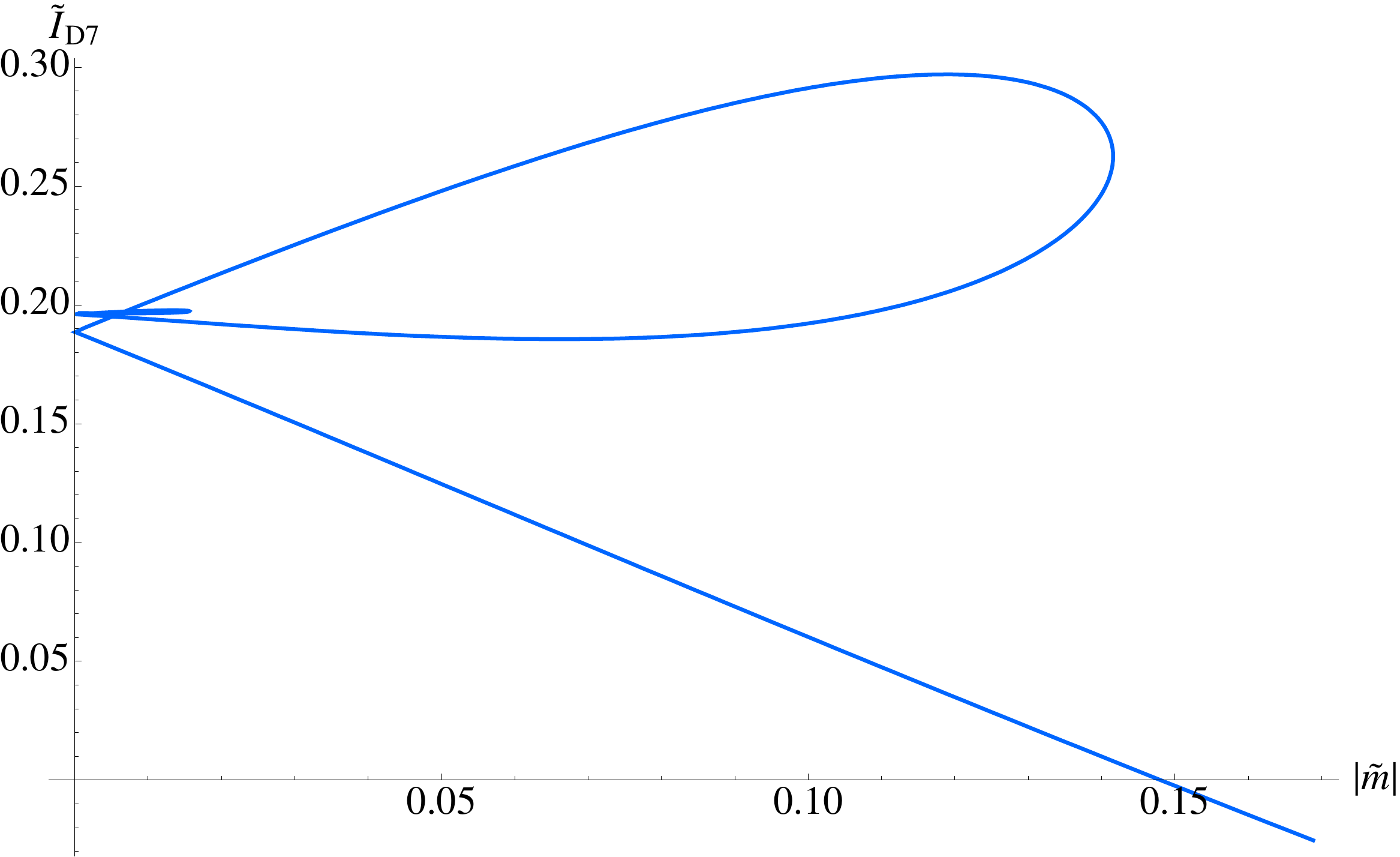}
   \caption{\small  The lowest lying curve correspond to the positive $\tilde m$ part of the lowest branch of the spiral, suggesting that this is the stable phase of the theory. }
   \label{fig:free-energy}
\end{figure}

\begin{figure}[h] 
   \centering
   \includegraphics[width=9cm]{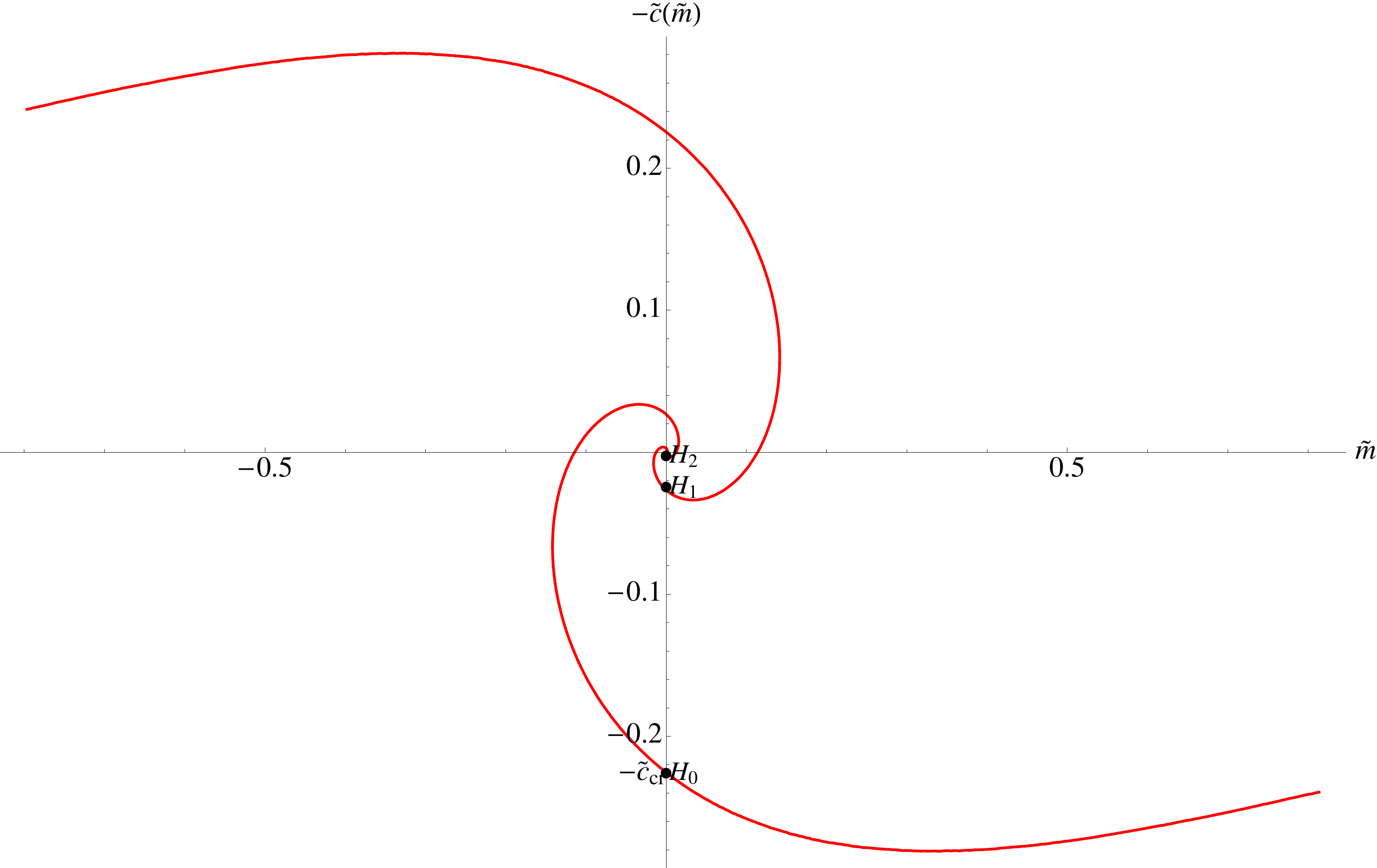}
   \caption{\small A magnification of figure \ref{fig:fig1} to show the spiral behavior near the origin of the $(-\tilde c,\tilde m)$-plane. We have added the second (left) arm of the spiral representing the $(\tilde m, -\tilde c)\to (-\tilde m,\tilde c)$ symmetry of the diagram.}
   \label{fig:spiral-revisited}
\end{figure}

 Another interesting feature of our
   phase diagram is the spiral behavior near the origin of the $(\tilde m,-\tilde c)$-plane which can be seen in figure \ref{fig:spiral-revisited}. Note that the spiral presented in this figure has two arms, we have used the fact that any two points in the $(\tilde m,-\tilde c)$ plane related by reflection with respect to the origin describe the same physical state. A similar spiraling feature has been observed
  in ref.~\cite{Albash:2006bs}, where the authors have argued that only the lowest branch of the spiral corresponding to positive values of
   $m$ is the stable one (corresponding to the lowest energy state).  The spiral behavior near the origin signals instability of the
   embedding corresponding to $L_0\equiv 0$. If we trace the curve of the diagram in
   figure \ref {fig:spiral-revisited} starting from large $m$, as we go to smaller values of $m$ we will reach zero bare quark mass for some
   large negative value of the fermionic condensate $c_{cr}$. Now if we continue tracing along  the diagram one can verify numerically that all other points correspond to embeddings of the D7-brane which intersect the origin of the transverse plane at least once. After further study of the right arm of the spiral, one finds  that the part of the diagram corresponding to negative values of $\tilde m$ represents solutions for the D7-brane embedding which intersect the origin of the transverse plane odd number of times, while the positive part of the spiral represents solutions which intersect the origin of the transverse plane even number of times. The lowest positive branch corresponds to solutions which don't intersect the origin of the transverse plane and is the stable one, while the upper branches have correspondingly $2,4, {\it etc.,}$ intersection points and are ruled out after evaluation of the free energy. 
   Indeed let us explore the stability of the spiral by calculating the regularized free energy of the system. We identify the free energy of the dual gauge theory \cite{Erdmenger:2007bn}, \cite{Albash:2007bk} with the wick rotated and regularized on-shell action of the D7-brane:
\begin{eqnarray}
&&F=2\pi^2N_fT_{D7}R^4H^2\tilde I_{D7}\ ,\\
&&\tilde I_{D7}=\int\limits_{0}^{\tilde\rho_{max}}d\tilde\rho\left({\tilde\rho}^3\sqrt{1+\frac{1}{({\tilde\rho}^2+{\tilde L}^2)}}\sqrt{1+{\tilde L}'^2}-\tilde\rho\sqrt{{\tilde\rho}^4+1}\right)\label{freeenergy}
\end{eqnarray}
The second term under the sign of the integral in (\ref{freeenergy}), corresponds to the subtracted free energy of the $\tilde L(\tilde\rho)\equiv 0$ embedding and serves as a regulator. 
Now we can evaluate numerically the integral in (\ref{freeenergy}) for the first several branches of the spiral. The corresponding plot is presented in figure \ref{fig:free-energy}. Note that we have plotted $\tilde I_{D7}$ versus $|\tilde m |$, since the bare quark mass depends only on the absolute value of the parameter $\tilde m$. The lowest curve on the plot corresponds to the lowest positive branch of the spiral, as one can see it has the lowest energy and thus corresponds to the stable phase of the theory.  
   
In the next section we will provide more detailed analysis of the spiral structure from Figure \ref{fig:spiral-revisited} and explore the discrete self-similarity associated to it. 

\section{Criticality and Spontaneous chiral symmetry breaking}
\subsection{The Spiral Revisited}
In the following section we analyze the spiral structure described in \cite{Filev:2007gb}. The technique that we employ is similar to the one used in \cite{Frolov:2006tc} and \cite{Mateos:2006nu} , where the authors studied merger transitions in brane-black-hole systems. 

Let us explore the asymptotic form of the equation of motion of the D7-brane probe (\ref{eqnMnLD}) in the near horizon limit $\tilde \rho^2+\tilde L^2\to 0$. To this end we change coordinates to:
\begin{eqnarray}
\tilde\rho\to\lambda\hat\rho;~~~ \tilde L\to \lambda\hat L;
\label{rescaling}
\end{eqnarray}
and consider the limit $\lambda\to0$. The resulting equation of motion is:
\begin{equation}
\partial_{\hat\rho}(\frac{\hat\rho^3}{\hat\rho^2+\hat L^2}\frac{\hat L'}{\sqrt{1+\hat L'^2}})+2\sqrt{1+\hat L'^2}\frac{\hat\rho^3\hat L}{(\hat\rho^2+\hat L^2)^2}=0\ .
\label{rescaled}
\end{equation} 
Equation (\ref{rescaled}) enjoys the scaling symmetry: 
\begin{equation}
\hat \rho\to \mu\hat\rho;~~~\hat L\to \mu\hat L;\ .
\end{equation}
In the sense that if $\hat L=f(\hat\rho)$ is a solution to the E.O.M. then $\frac{1}{\mu}f(\mu\hat\rho)$ is also a solution.
Next we focus on the region of the parametric space, close to the trivial $L\equiv 0$ embedding, by considering the expansion:
\begin{equation}
\hat L=0+(2\pi\alpha')\hat\chi
\end{equation}
and linearizing the E.O.M. . The resulting equation of motion is:
\begin{equation}
\hat\rho\partial_{\hat\rho}(\hat\rho\partial_{\hat\rho}\hat\chi)+2\hat\chi=0
\end{equation}
and has the solution :
\begin{equation}
\hat\chi=A\cos(\sqrt{2}\ln\hat\rho)+B\sin(\sqrt{2}\ln\hat\rho)\ .
\label{linafter}
\end{equation}
Now under the scaling symmetry $\hat\rho\to\mu\hat\rho$ the constants of integration $A$ and $B$ transform as:
\begin{equation}
\begin{pmatrix}A \\ B\\ \end{pmatrix}\to\frac{1}{\mu}\begin{pmatrix}\cos\sqrt{2}\ln\mu  &\sin\sqrt{2}\ln\mu \\ -\sin\sqrt{2}\ln\mu &\cos\sqrt{2}\ln\mu \end{pmatrix}\begin{pmatrix}A\\ B \end{pmatrix}\ .
\label{scaling1}
\end{equation}
The above transformaton defines a class of solutions represented by a logarithmic spiral in the parametric space $(A,B)$ generated by some $(A_{in},B_{in})$, the fact that we have a discrete symmetry $\chi\to-\chi$ suggests that $(-A_{in},-B_{in})$ is also a solution and therefore the curve of solutions in the parametric space is a double spiral symmetric with respect to the origin. Actually as we are going to show there is a linear map from the parametric space   $(A,B)$ to the plane $(\tilde m,-\tilde c)$ which explains the spiral structure, a subject of our study.
To show this let us consider the linearized E.O.M. before taking the $\lambda\to 0$ limit :
\begin{eqnarray}
\tilde\rho\sqrt{1+\tilde\rho^4}\partial_{\tilde\rho}(\tilde\rho\sqrt{1+\tilde\rho^4}\partial_{\tilde\rho}\tilde\chi)+2\tilde\chi=0;~~~
\tilde\chi=\lambda\hat\chi;\ ,
\end{eqnarray}
with the solution:
\begin{equation}
\tilde\chi=\tilde A\cos\sqrt{2}\ln\frac{\tilde\rho}{\sqrt{1+\sqrt{1+\tilde\rho^4}}}+\tilde B\sin\sqrt{2}\ln\frac{\tilde\rho}{\sqrt{1+\sqrt{1+\tilde\rho^4}}}\ .
\label{linbefore}
\end{equation}
Expanding at infinity:
\begin{eqnarray}
\tilde\chi=\tilde m+\frac{\tilde c}{\tilde\rho^2}+\dots=\tilde A-\frac{\tilde B}{\sqrt{2}}\frac{1}{\tilde\rho^2}+\dots,
\end{eqnarray}
we get:
\begin{equation}
\begin{pmatrix}\tilde m \\ \tilde c\end{pmatrix}=\begin{pmatrix}\tilde A\\-{\tilde B}/{\sqrt{2}}\end{pmatrix}\ .
\end{equation}
Now if we match our solution (\ref{linbefore}) with the solution in the $\tilde\rho\to 0$ limit (\ref{linafter}) we should identify $(\tilde A,\tilde B)$ with the parameters $(A, B)$. Combining the rescaling property of $(A,B)$ with the linear map to $(\tilde m,-\tilde c)$ we get that the embeddings close to the trivial embedding $L\equiv 0$ are represented in the $(\tilde m,-\tilde c)$ plane by a double spiral defined {\it via} the transformation:
\begin{equation}
\begin{pmatrix}\tilde m\\ \tilde c\\ \end{pmatrix}\to\frac{1}{\mu}\begin{pmatrix}\cos\sqrt{2}\ln\mu &-\sqrt{2}\sin\sqrt{2}\ln\mu \\ \frac{1}{\sqrt{2}}\sin\sqrt{2}\ln\mu & \cos\sqrt{2}\ln\mu \end{pmatrix}\begin{pmatrix}\tilde m\\ \tilde c \end{pmatrix}\ .
\label{scaling2}
\end{equation}
Note that the spiral is double, because we have the symmetry $(\tilde m,-\tilde c)\to(-\tilde m,\tilde c)$. This implies that in order to have similar configurations at scales $\mu_{1}$ and $\mu_{2}$ we should have:
\begin{equation}
\cos\sqrt{2}\ln\mu_1=\pm\cos\sqrt{2}\ln\mu_2
\end{equation}
and hence :
\begin{equation}
\sqrt{2}\ln\frac{\mu_2}{\mu_{1}}=-n\pi,
\end{equation}
which is equivalent to:
\begin{equation}
\frac{\mu_2}{\mu_1}=e^{-n\pi/\sqrt{2}}=q^n\ .
\end{equation}
Therefore we obtain that the discrete self-similarity is described by a rescaling by a factor of:
\begin{equation}
q=e^{-\pi/\sqrt{2}}\approx 0.10845\ .
\label{q}
\end{equation}
This number will appear in the next subsection where we will study the meson spectrum. As one may expect the meson spectrum also has a self-similar structure.  
 
 It is interesting to confirm numerically the self-similar structure of the spiral and to calculate the critical exponents of the bare quark mass and the fermionic condensate. It is convenient to use the separation of the D3 and D7 branes at $\tilde\rho=0$, $\tilde L_{in}=\tilde L(0)$ as an order parameter.  There is a discrete set of initial separations $L_{in}$, corresponding to the points $H_0, H_1, H_2, \dots$ in figure \ref{fig:spiral-revisited} , for which the corresponding D7 brane's embeddings asymptote to $\tilde m=\tilde L_{\infty}=0$ as $\tilde\rho\to\infty$. The trivial $\tilde L\equiv 0$ embedding has ${\tilde L}_{in}=0$ and is the only one which has a zero fermionic condensate $(\tilde c=0)$, the rest of the states have a non zero $\tilde c$ and hence a chiral symmetry is spontaneously broken. Each such point determines separate branch of the spiral where $\tilde c=\tilde c(\tilde m)$ is a single valued function. On the other side each such branch has both positive $\tilde m$ and negative $\tilde m$ parts. The symmetry of the double spiral from figure \ref{fig:spiral-revisited}, suggests that the states with negative $\tilde m$ are equivalent to positive $\tilde  m$ states but with an opposite sign of  $\tilde c$. This implies that the positive and negative $\tilde m$ parts of each branch correspond to two different phases of the theory, with opposite signs of the condensate. As we can see from figure \ref{fig:free-energy} the lowest positive branch of the spiral has the lowest free energy and thus corresponds to the stable phase of the theory. In the next subsection we will analyze the stability of the spiral further by studying the light meson spectrum of the theory near the critical $\tilde L\equiv 0$ embedding.
 
\begin{figure}[p] 
   \centering
   \includegraphics[width=11cm]{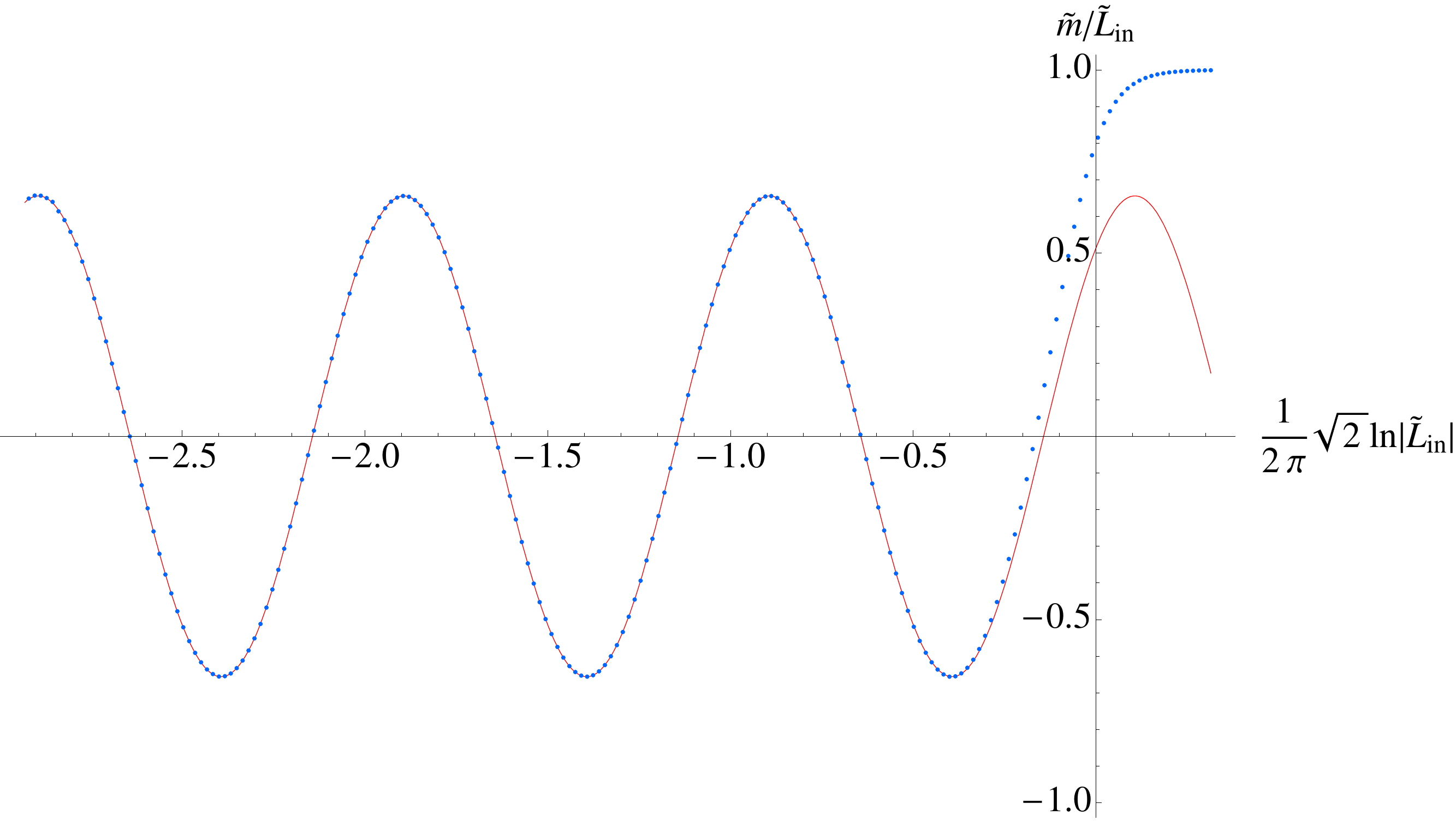} 
  
   \includegraphics[width=11cm]{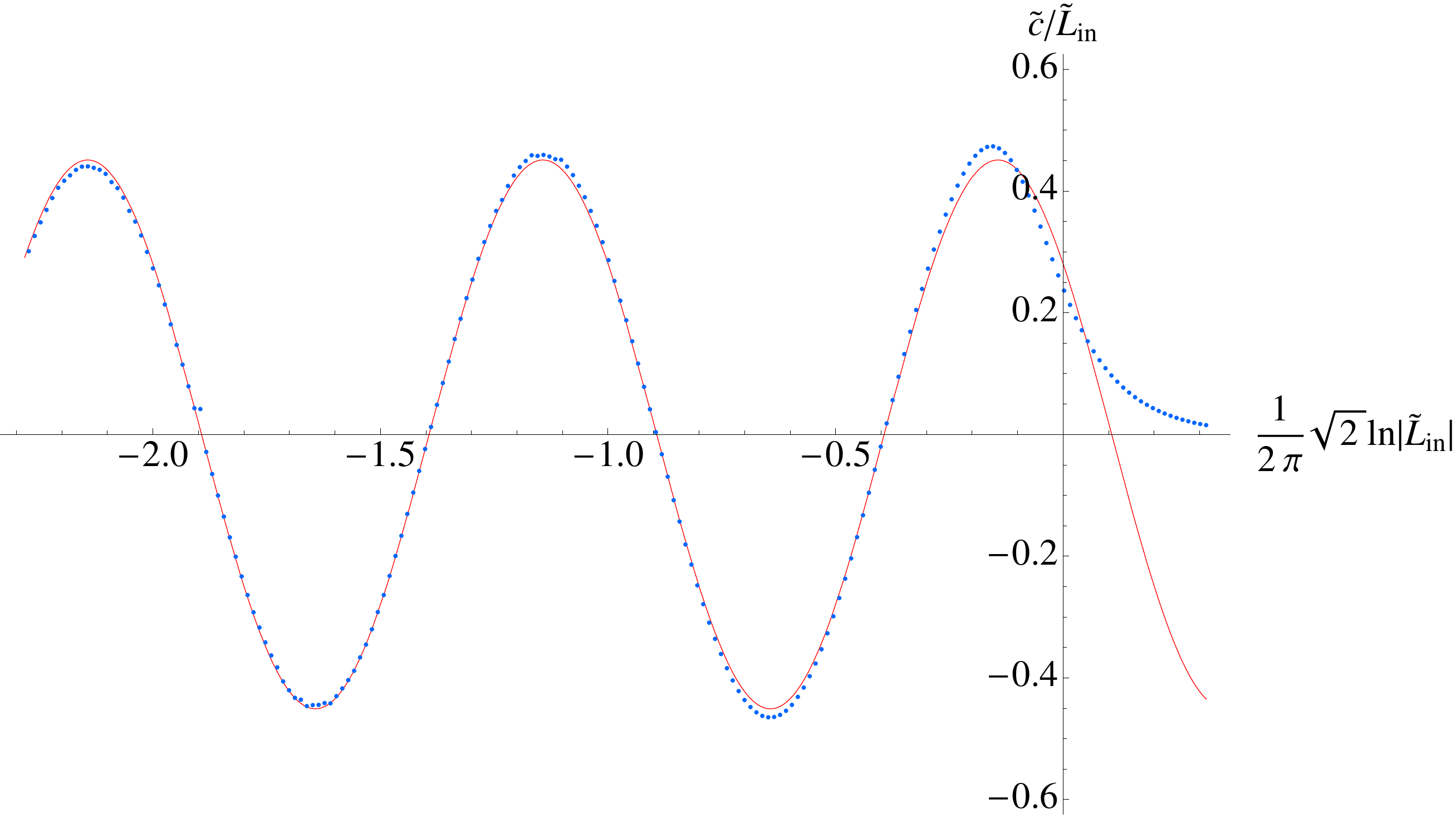} 
   \caption{\small The red curves represent fit with trigonometric functions of unit period. For small $\tilde L_{in}$ the fit is very good, while for large $\tilde L_{in}$ we get the results for pure $AdS_{5}\times S^5$ space, namely $\tilde L=const$, $\tilde c=0$. The plots also verify that the critical exponents of $\tilde m$ and $\tilde c$ are equal to one.}
   \label{fig:mspiral}
\end{figure}

 Here we are going to show that both the bare quark mass $\tilde m$ and the fermionic condensate $\tilde c$ have critical exponent one, as $\tilde L_{in} \to 0$. Indeed let us consider the scaling property (\ref{scaling1}), (\ref{scaling2}). If we start from some $\tilde L_{in}^0$ and transform to $\tilde L_{in}=\frac{1}{\mu} \tilde L_{in}^0$, we can solve for $\mu$ and using equation (\ref{scaling2}) we can verify that the bare quark mass and the fermionic condensate approach zero linearly as $\tilde L_{in}\to0$. To verify numerically our analysis we generated plots of $\tilde m/\tilde L_{in}$ vs. $\sqrt{2}\log{\tilde L_{in}}/2\pi$  and $\tilde c/\tilde L_{in}$ vs. $\sqrt{2}\log{\tilde L_{in}}/2\pi$ presented in figure \ref{fig:mspiral}.
 
The red curves in these figures represent a fit with trigonometric functions of a unit period, as one can see the fit is very good as $\tilde L_{in}\to 0$. On the other side for large $\tilde L_{in}$ we obtain the results for a pure $AdS_{5}\times S^5$ space, namely $\tilde L=const$, $\tilde c=0$. It is also evident from the plots that the critical exponents of $\tilde m$ and $\tilde c$ are equal to one.

\subsection{The Meson Spectrum}
In this section we will explore the light meson spectrum of the theory corresponding to quadratic fluctuations of the D7 brane embedding. In particular we will consider the spectrum corresponding to the fluctuations of $\tilde L$. The equations of motion of the fluctuation modes were derived in \cite{Filev:2007gb} and it was shown that the vector and the scalar spectrum mix due to the non-zero magnetic field. Some interesting effects such as Zeeman splitting of the states and a characteristic $\sqrt{m}$ dependence of the meson spectrum have been reported. However the analysis performed in  \cite{Filev:2007gb} is only for the fluctuations along $\phi$, for the lowest positive branch of the spiral from figure \ref{fig:spiral-revisited} (the one corresponding to point $H_0$). In this letter we extend the analysis of the spectrum to all branches of the spiral (points $H_1, H_2,\dots$ in figure \ref{fig:spiral-revisited}) and  show that the ground states of all inner branches of the spiral are tachyonic, proving that the phases described by these branches of the spiral are unstable as opposed to metastable. Our analysis reveals the self-similar structure of the spectrum and we obtain the critical exponents of the tachyonic spectrum as one approaches the critical $\tilde L\equiv 0$ embedding. The chapter is organized as follows: 

First we study the spectrum of the $\tilde L\equiv 0$ embedding in the spirit of the analysis provided in \cite{Hoyos:2006gb}. We perform both a numerical and analytical study and show that the spectrum contains infinitely many tachyonic states approaching zero in a decreasing geometrical series, representing the self-similar structure of the meson spectrum.

Next we study the spectrum as a function of the bare quark mass and show that at each turn of the spiral one of the energy levels become tachyonic. Similar behavior has been recently reported in \cite{Mateos:2007vn}. We show that as we approach the critical $\tilde L\equiv 0$ embedding the spectrum becomes tachyonic and the corresponding critical exponent is two. We also present plots showing the spiraling of the spectrum as one approaches criticality.

Finally we provide an analysis of the spectrum of the stable branch of the spiral and comment on the small $\tilde m$ behavior of the spectrum as a consistent with the spontaneous chiral symmetry breaking scenario.

\subsubsection{The critical $\tilde L\equiv 0$ embedding} 

In this section we study the $\tilde L\equiv 0$ embedding and in particular the spectrum of the fluctuations along the $\tilde L$ coordinate. Let us go back to dimensionfull coordinates and consider the following change of coordinates in the transverse $R^6$ space:

\begin{eqnarray}
\rho=u\cos\theta\ ,\\
L=u\sin\theta\nonumber\ .
\end{eqnarray}
In these coordinates the trivial embedding corresponds to $\theta\equiv 0$ and in order to study the quadratic fluctuations we perform the expansion:
\begin{eqnarray}
\theta=0+(2\pi\alpha')\delta\theta(t,u)\ ,\\
\delta\theta=e^{-i\Omega t}h(u)\ .
\end{eqnarray}

Note that in order to study the mass spectrum we restrict the D7 brane to fluctuate only in time. In a sense this corresponds to going to the rest frame. Note that due to the presence of the magnetic field there is a coupling of the scalar spectrum to the vector one, however for the fluctuations along $\theta$ the coupling depends on the momenta in the $(x_2,x_3)$ plane and this is why considering the rest frame is particularly convenient . 
 
\begin{figure}[h] 
   \centering
   \includegraphics[width=10cm]{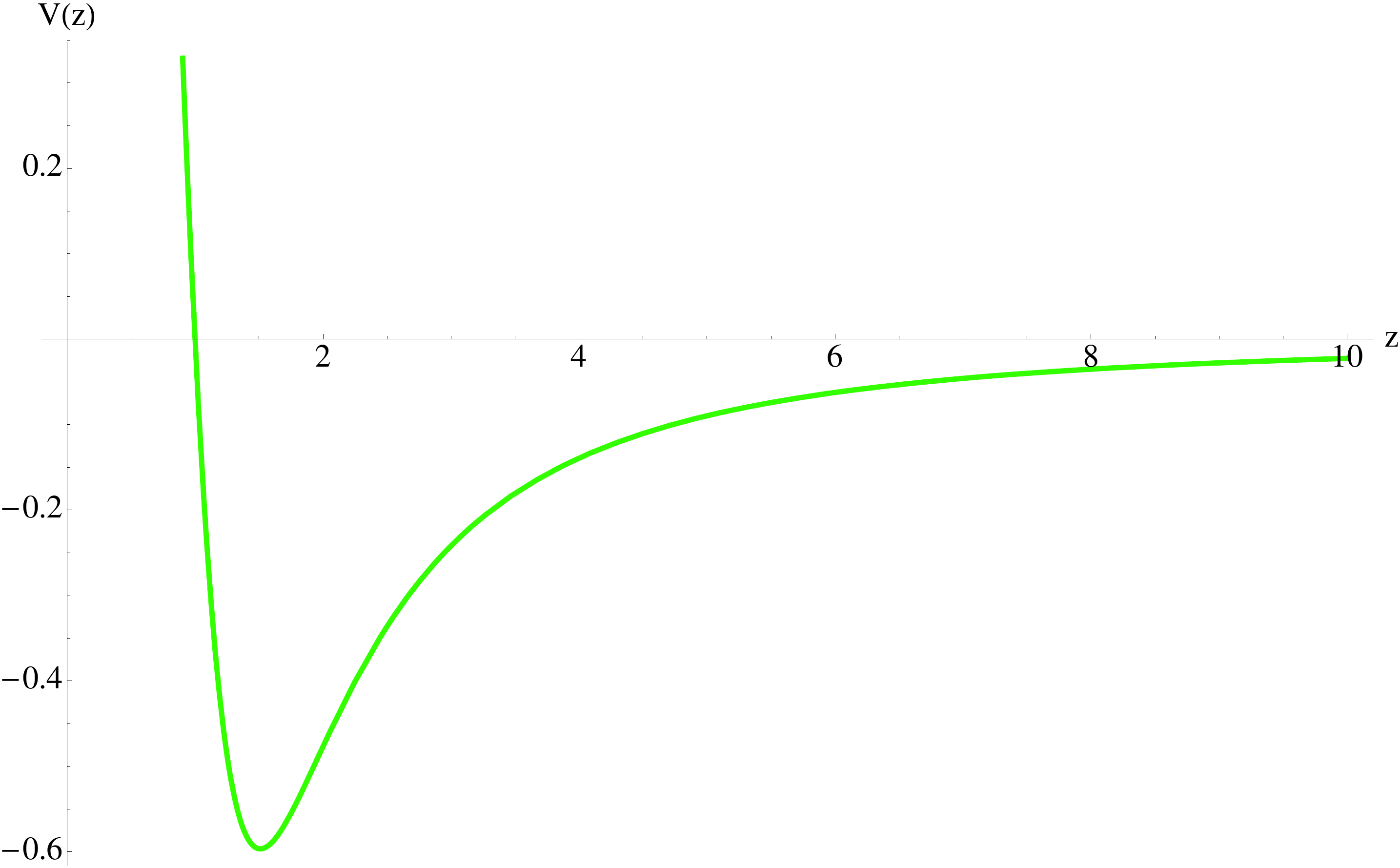} 
   \caption{\small A plot of the effective potential $V(z)$ given in equation (\ref{Potential}). }
   \label{fig:Potential}
\end{figure}

Our analysis follows closely the one considered in \cite{Hoyos:2006gb}, where the authors have calculated the quasinormal modes of the D7-brane embedding in the AdS-black hole background by imposing an in-going boundary condition at the horizon of the black hole. Our case is the $T\to0$ limit and the horizon is extremal, however the $\theta\equiv 0$ embedding can still have quasinormal excitations  with imaginary frequencies, corresponding to a real wave function so that there is no flux of particles falling into the zero temperature horizon.
The resulting equation of motion is:
\begin{equation}
h''+\left(\frac{3}{u}+\frac{2 u^3}{u^4+R^4H^2}\right)h'+\left(\frac{R^4}{u^4}\omega^2+\frac{3}{u^2}\right)h=0\ .
\end{equation}
It is convenient to introduce the following dimensionless quantities:
\begin{equation}
z=\frac{R}{u}\sqrt{H};~~~\omega=\frac{\Omega R}{\sqrt{H}};\ ,
\end{equation}
and make the substitution \cite{Hoyos:2006gb}
\begin{equation}
h(z)=\sigma(z)f(z);~~~\frac{\sigma'(z)}{\sigma(z)}=\frac{1}{2z}+\frac{1}{z(1+z^4)};\ ,
\end{equation}
leading to the equation for the new variable $f(z)$:
\begin{equation}
f''(z)+\left(\omega^2-V(z)\right)f(z)=0\ .
\label{Shr}
\end{equation}
Where the effective potential is equal to:
\begin{equation}
V(z)=\frac{3}{4z^2}\frac{(1+3z^4)(1-z^4)}{(1+z^4)^2}\ .
\label{Potential}
\end{equation}
The potential in (\ref{Potential}) goes as $\frac{3}{4z^2}$ for $z\to 0$ and as $-\frac{9}{4z^2}$ for $z\to\infty$ and is presented in figure \ref{fig:Potential}. As it was discussed in \cite{Hoyos:2006gb} if the potential gets negative the imaginary part of the frequency may become negative. Furthermore the shape of the potential suggests that there might be bound states with a negative $\omega^2$. To obtain the spectrum we look for regular solutions of  (\ref{Shr}) imposing an in-falling boundary condition at the horizon ($z\to\infty$). 

The asymptotic form of the equation of motion at $z\to\infty$ is that of the harmonic oscillator:
\begin{equation}
f''(z)+\omega^2f(z)=0\ ,
\end{equation}
with the solutions $e^{\pm i\omega z}$, the in-falling boundary condition implies that we should choose the positive sign. In our case the corresponding spectrum turns out to be tachyonic and hence the exponents are real. Therefore the in-falling boundary condition simply means that we have selected the regular solution at the horizon: $z\to \infty$. We look for a solution of the form:
\begin{equation}
f(z)=e^{+i\omega z}S(z)\ .
\end{equation}
The resulting equation of motion for $S(z)$ is:
\begin{equation}
(-3-6z^4+9z^8)S(z)+4z^2(1+z^4)^2\left(2i\omega S'(z)+S''(z)\right)=0\ .
\label{eqnS}
\end{equation}

Next we study numerically equation (\ref{eqnS}). After solving the asymptotic form of the equation at the Horizon, we impose the following boundary condition at $z=1/\epsilon$, where $\epsilon$ is a numerically small number typically $\epsilon=10^{-9}$ :
\begin{equation}
S(1/\epsilon)=1-\frac{9i\epsilon}{8\omega};~~~S'(1/\epsilon)=\frac{9i\epsilon^2}{8\omega};\ ,
\label{initialcond}
\end{equation}
after that we explore the solution for a wide range of $\omega=i\omega_{I}$. We look for regular solutions which have $|S(\epsilon)|\approx 0$, this condition follows from the requirement that $\chi \propto z^3$ as $z\to 0$. It turns out that regular solutions exist for a discrete set of positive $\omega_{I}\ll1$. The result for the first six modes that we obtained is presented in table \ref{tab:1}.

    \begin{table}[h]
    \caption{}
\begin{center}
\begin{tabular}{|c|c|c|}
\hline
 $n$&$\omega_{I}^{(n)}$&$\omega_{I}^{(n)}/\omega_{I}^{(n-1)}$\\\hline
 0&$2.6448\times10^{-1}$&-\\\hline
 1&$2.8902\times10^{-2}$&0.10928\\\hline
 2&$3.1348\times10^{-3}$&0.10846\\\hline
 3&$3.3995\times10^{-4}$&0.10845\\\hline
 4&$3.6865\times10^{-5}$&0.10844\\\hline
 5&$3.9967\times10^{-6}$&0.10841\\\hline
\end{tabular}
\end{center}
\label{tab:1}
\end{table}%

\begin{figure}[h] 
   \centering
   \includegraphics[width=12cm]{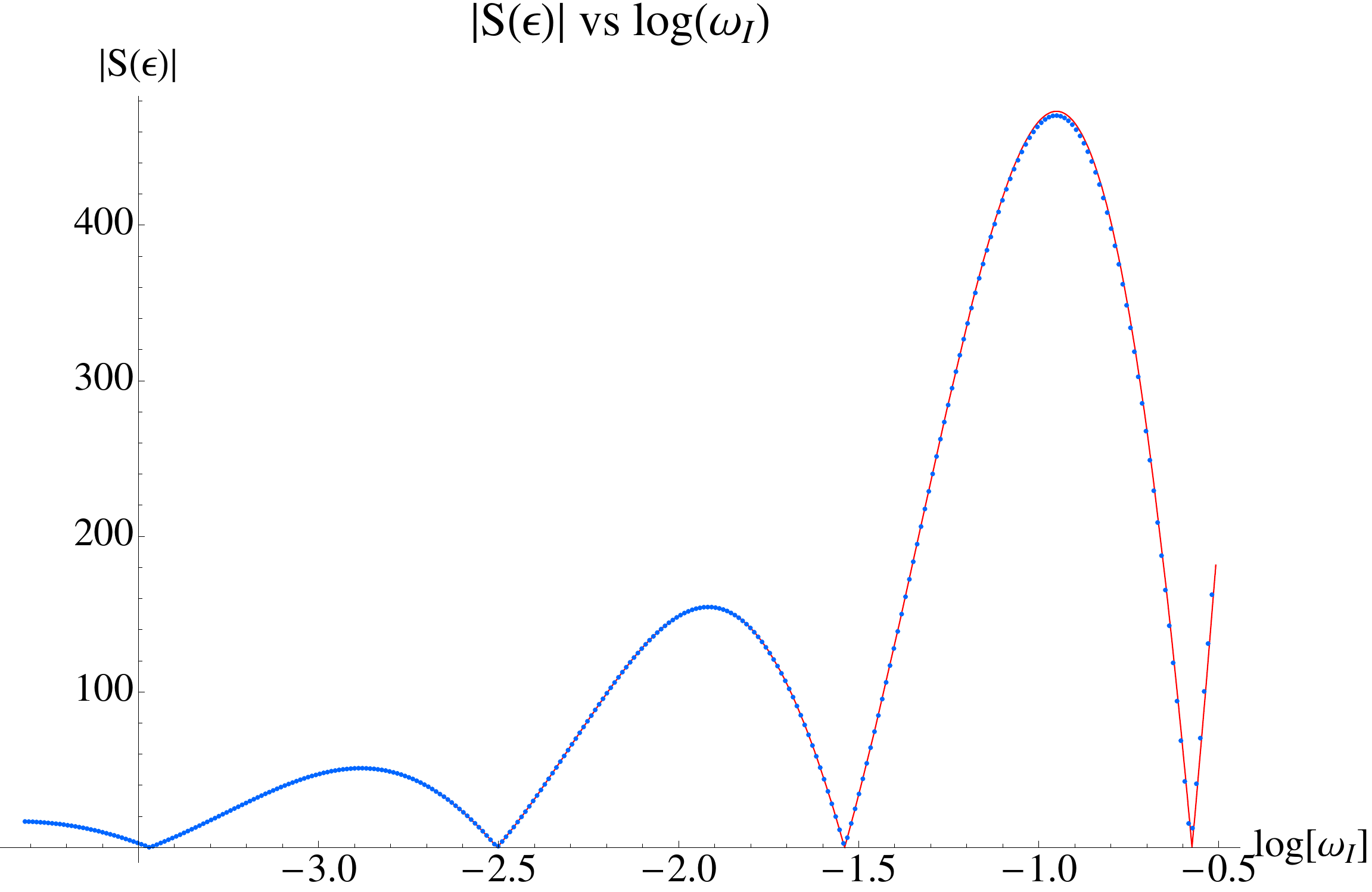} 
   \caption{\small The dotted blue curve corresponds to the numerical solution of equation (\ref{eqnS}), while the thick red curve is the one determined by equation (\ref{Shat2}). The plots are scaled to match along the vertical axis.}
   \label{fig:spectrum}
\end{figure}

The data suggests that as $\omega_{I}\to 0$ the states organize in a decreasing geometrical series with a factor $q\approx0.1084$. Up to four significant digits, this is the number from equation (\ref{q}), which determines the period of the spiral. We can show this analytically. To this end let us consider the rescaling of the variables in equation (\ref{eqnS}) given by:
\begin{eqnarray}
z=\lambda \hat z;~~~\hat\omega=\omega/\lambda;~~~\lambda\to\infty;\ .
\end{eqnarray}
This is leading to:
\begin{equation}
9\hat S(\hat z)+4\hat z^2(2i\hat\omega\hat S'(\hat z)+\hat S''(\hat z))+O({\lambda}^{-4})=0\ .
\label{small}
\end{equation}
The solution consistent with the initial conditions at infinity  (\ref{initialcond}) can be found to be:
\begin{equation}
\hat S(\hat z)=\frac{1+i}{2} e^{-i\frac{\pi}{\sqrt{2}}}e^{-i\hat z\hat\omega}\sqrt{\pi\hat z\hat\omega}H_{i\sqrt{2}}^{(1)}(\hat z\hat\omega);~~~\hat\omega=i\hat\omega_{I};\ ,
\label{Shat}
\end{equation}
where $H_{i\sqrt{2}}^{(1)}$ is the Hankel function of the first kind. Our next assumption is that in the $\omega_I\to 0 $ limit, this asymptotic form of the equation describes well enough the spectrum. To quantize the spectrum we consider some $\hat z_0=z_0/\lambda\ll1$, where we have $1\ll z_0\ll\lambda$ so that the simplified form of equation (\ref{small}) is applicable and impose:
\begin{equation}
\hat S(\hat z_0)=0\ .
\label{Shat2}
\end{equation}
Using that $\hat z\hat\omega=iz\omega_I$ this boils down to:
\begin{equation}
H_{i\sqrt{2}}^{(1)}(i\omega_I z_0)=0\ .
\end{equation}
Now using that $\omega_I z_0\ll1$ for a sufficiently small $\omega_I$, we can make the expansion:
\begin{equation}
H_{i\sqrt{2}}^{(1)}(i\omega_I z_0)\approx-A_1\left((\omega_I z_0)^{i\sqrt{2}}-(\omega_I z_0)^{-i\sqrt{2}}\right)+iA_2\left((\omega_I z_0)^{i\sqrt{2}}+(\omega_I z_0)^{-i\sqrt{2}}\right)\ ,
\end{equation}
where $A_1$ and $A_2$ are real numbers defined {\it via}:
\begin{equation}
A_1+iA_2=-\frac{1}{\pi}{i(i/2)^{-i\sqrt{2}}\Gamma(i\sqrt{2})}\ .
\end{equation}
This boils down to:
\begin{equation}
\cos(\sqrt{2}\ln(\omega_I z_0)+\phi)=0;~~~\phi\equiv\pi/2-\arg(A_1+iA_2);\ .
\label{20}
\end{equation}
The first equation in (\ref{20}) leads to:
\begin{equation}
\omega_I^{(n)}=\frac{1}{z_0}e^{-\frac{\pi/2+\phi}{\sqrt{2}}}e^{-n\frac{\pi}{\sqrt{2}}}=\omega_I^{(0)}q^n\ ,
\label{geom}
\end{equation}
suggesting that:
\begin{equation}
q=e^{-\frac{\pi}{\sqrt{2}}}\approx 0.10845\ .
\label{an_q}
\end{equation}
This is the number given in (\ref{q}). Note that the value of $z_0$ is a free parameter that we can fix by matching equation (\ref{geom}) to the data in table \ref{tab:1}. On the other side $\hat S(\hat z)$ given in equation (\ref{Shat}) depends only on $\hat z\hat\omega=i\omega_I z$ and therefore once we have fixed $z_0$ we are left with a function of $\omega_I$, which zeroes determine the spectrum, equation (\ref{Shat2}). It is interesting to compare it to the numerically obtained plot of $|S(\epsilon)|$ vs. $\omega_I$, that we have used to determine the spectrum numerically. The result is presented in figure \ref{fig:spectrum}, where we have used the $n=3$ entry from table \ref{tab:1} to fix $z_0$. One can see the good agreement between the spectrum determined by equation (\ref{Shat2}), the red curve in figure \ref{fig:spectrum} and the numerically determined one, the dotted blue curve.

\subsubsection{The Spectrum near criticality}
In this chapter we study the light meson spectrum of the states forming the spiral structure in the $(\tilde m, -\tilde c)$ plane, figure \ref{fig:spiral-revisited}. In particular we focus on the study of the fluctuations along $ L$. The corresponding equation of motion was derived in \cite{Filev:2007gb}. The effect of the magnetic field $H$ is to mix the vector and the meson parts of the spectrum. However if we consider the rest frame by allowing the fluctuations to depend only on the time direction of the D3 branes' world volume, the equation of motion for the fluctuations along $ L$ decouple from the vector spectrum. To this end we expand:
\begin{eqnarray}
L=L_0(\rho)+(2\pi\alpha')\chi(\rho,t)\ ,\\
\chi=h(\rho)\cos{M t}\nonumber\ .
\end{eqnarray}
Here $L_0(\rho)$ is the profile of the D7 brane's classical embedding. The resulting equation of motion for $h(\rho)$ is:
\begin{eqnarray}
&\partial_\rho(g\frac{h'}{(1+L_0'^2)^2})+\left(g\frac{R^4}{(\rho^2+L_0^2)^2}\frac{M^2}{1+L_0'^2}-\frac{\partial^2 g}{\partial L_0^2}+\partial_\rho(\frac{\partial g}{\partial L_0}\frac{L_0'}{1+L_0'^2})\right)h=0\ ,\\
{\rm where}\quad&g(\rho,L_0,L_0')=\rho^3\sqrt{1+{L_0}'^2}\sqrt{1+\frac{R^4H^2}{(\rho^2+L_0^2)^2}}\ .\nonumber
\end{eqnarray}
It is convenient to introduce the dimensionless variables:
\begin{equation}
\tilde h=\frac{h}{R\sqrt{H}};~~\tilde L_{0}=\frac{L_0}{R\sqrt{H}};~ \tilde\rho=\frac{\rho}{R\sqrt{H}};~\tilde M=\frac{M R}{\sqrt{H}};\ ,
\label{dimensionless}
\end{equation}
leading to:
\begin{eqnarray}
&\partial_{\tilde\rho}(\tilde g\frac{\tilde h'}{(1+\tilde L_0'^2)^2})+\left(\tilde g\frac{1}{(\tilde\rho^2+\tilde L_0^2)^2}\frac{\tilde M^2}{1+\tilde L_0'^2}-\frac{\partial^2 \tilde g}{\partial \tilde L_0^2}+\partial_{\tilde\rho}(\frac{\partial \tilde g}{\partial \tilde L_0}\frac{\tilde L_0'}{1+\tilde L_0'^2})\right)\tilde h=0\ ,\label{fluctPsi}\\
{\rm with}\quad&\tilde g(\tilde\rho,\tilde L_0,\tilde L_0')=\tilde\rho^3\sqrt{1+{\tilde L_0}'^2}\sqrt{1+\frac{1}{(\tilde\rho^2+\tilde L_0^2)^2}}\nonumber\ .
\end{eqnarray}

We study the normal modes of the D7 brane described by equation (\ref{fluctPsi}) by imposing Neumann boundary conditions at $\tilde\rho=0$. Since our analysis is numerical we solve the equation of motion (\ref{fluctPsi}) in terms of a power series for small $\tilde\rho$ and impose the appropriate initial conditions for the numerical solution at $\tilde\rho=\epsilon$, where $\epsilon$ is some very small number. In order to quantize the spectrum we look for numerical solutions which are normalizable and go as $1/\tilde\rho^2$ at infinity. 
 
\begin{figure}[htbp] 
   \centering
   \includegraphics[width=12cm]{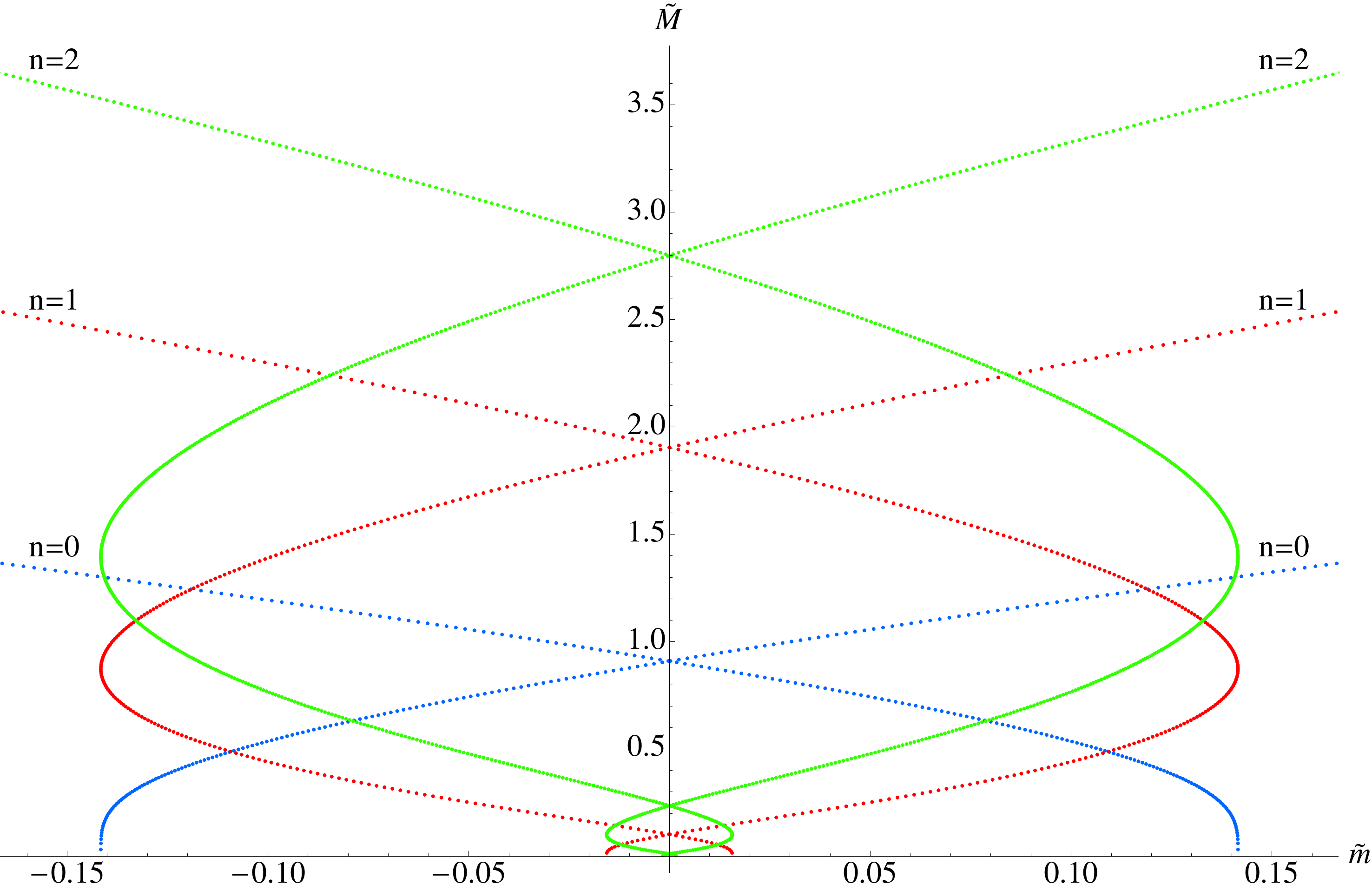} 
     \includegraphics[width=12cm]{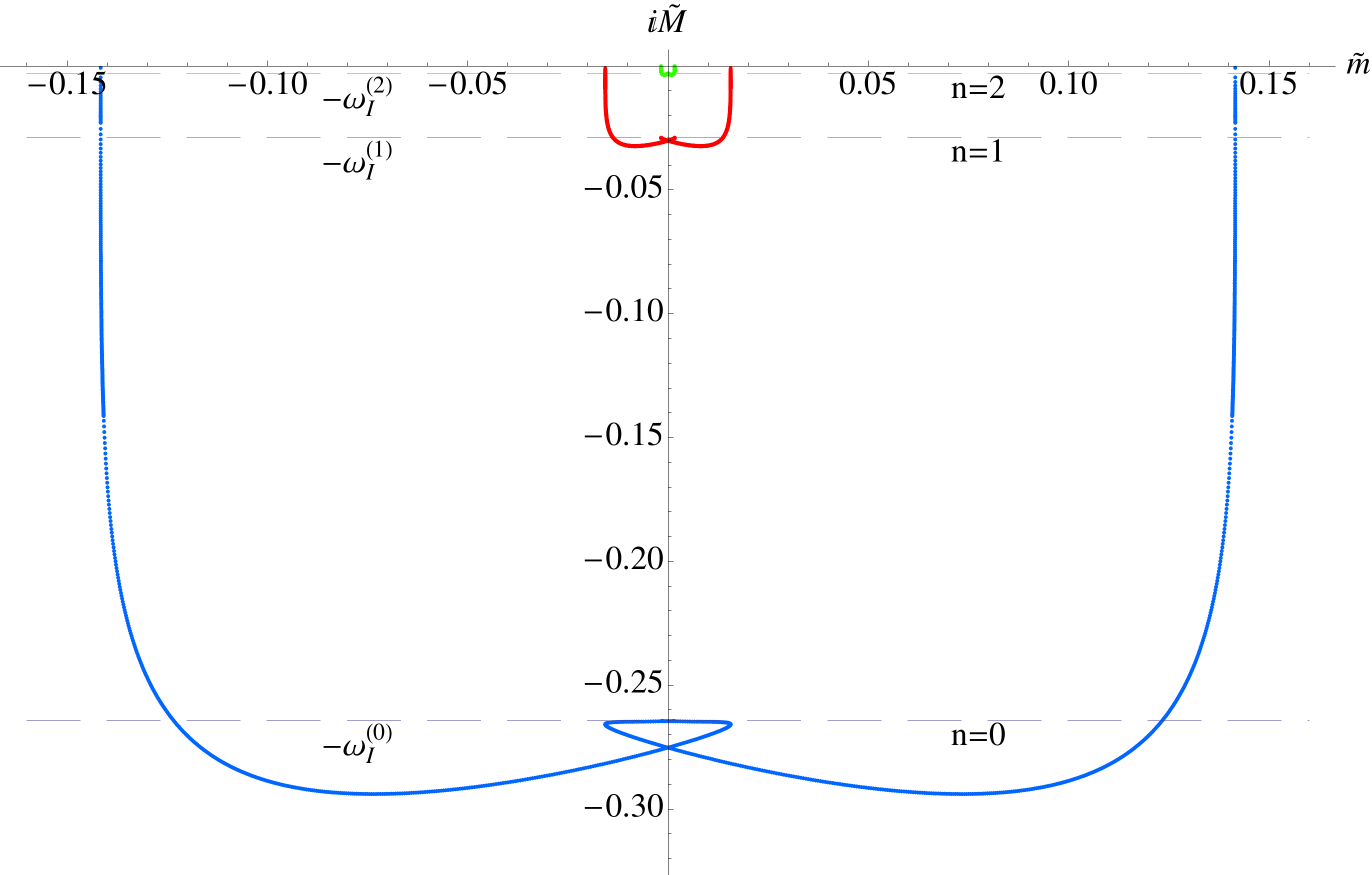} 
   \caption{\small A plot of the meson spectrum corresponding to the two arms of the spiral structure at the origin of the $(\tilde m,-\tilde c)$ plane. The ground state ($n=0$) becomes tachyonic for the inner branches of the spiral, while only the lowest branch is a tachyon free one. The tachyon sector of the diagram reveals the self-similar structure of the spectrum.}
\label{fig:spectrum-spiral}
\end{figure}

 Let us study the dependence of the spectrum of $\tilde M$ on the bare quark mass $\tilde m$, for the states corresponding to the spiral structure from figure \ref{fig:spiral-revisited}. A plot of the spectrum of the first three excited states is presented in figure \ref{fig:spectrum-spiral}. The classification of the states in terms of the quantum number $n$ is justified, because at large $\tilde m$ the equation of motion for the fluctuations asymptotes to the equation of motion for the pure $AdS_5\times S^5$ space, considered in \cite{Kruczenski:2003be}, where the authors obtained the spectrum in a closed form. Note that the diagram has a left-right symmetry. This is because we plotted the spectrum for both arms of the spiral in order to emphasize its self-similar structure, physically only one side of the diagram is sufficient. 
 
 Let us trace the blue curve corresponding to the $n=0$ state starting from the right-hand side. As $\tilde m$ decreases the mass of the meson decreases and at $\tilde m=0$ it has some non-zero value. This part of the diagram corresponds to the lowest positive branch of the spiral from figure \ref{fig:spiral-revisited} (the vicinity of point $H_0$). 
 It is satisfying to see that the lowest positive $\tilde m$ branch of the spiral is tachyon free and therefore stable under quantum fluctuations. Note that despite that the negative $\tilde m$ part of the lowest branch has no tachyonic modes in its fluctuations along $L$, it has a higher free energy (as can be seen from figure \ref{fig:free-energy}) and is thus at best metastable.   
 
 One can also see that the spectrum drops to a zero and becomes tachyonic exactly at the point where we start exploring the upper branch of the spiral. This proves that all inner branches correspond to true instability of the theory and cannot be reached by super-cooling. As we go deeper into the spiral, the $n=0$ spectrum remains tachyonic and spirals to some critical value. The dashed line denoted by $\omega_I^{(0)}$ in figure \ref{fig:spiral-revisited}  corresponds to the first entry in table \ref{tab:1}. As one can see this is the critical value approached by the spectrum. 
 
 Now let us comment on the $n=1,2$ levels of the spectrum represented by the red and green curves, respectively. As one can see the $n=1$ spectrum becomes tachyonic when we reach the third branch of the spiral (the vicinity of point $H_2$ in figure \ref{fig:spiral-revisited}) and after that follows the same pattern as the $n=0$ level, spiraling to the second entry $\omega_I^{(1)}$ from table \ref{tab:1}. The $n=2$ level has a similar behavior, but it becomes tachyonic at the next turn of the spiral and it approaches the next entry from table \ref{tab:1}. Similar feature was reported recently in \cite{Mateos:2007vn} where the authors  studied topology changing transitions. The above analysis suggests that at each turn of the spiral, there is one new tachyonic state appearing. It also suggests that the structure of the $n$-th level is similar to the structure of the $n+1$-th level and in the $n\to\infty$ limit this similarity becomes an exact discrete self-similarity. The last feature is apparent from the tachyonic sector of the diagram in the second plot in figure \ref{fig:spectrum-spiral}, the blue, red and green curves are related by an approximate scaling symmetry, the analysis of the spectrum of the critical $L\equiv 0$ embedding suggests that this symmetry becomes exact in the $n\to\infty$ limit with a scaling factor of $q$ given  in equation (\ref{q}).
 
 It is interesting to analyze the way the meson mass $\tilde M$ approaches its critical value and compute the corresponding critical exponent. Let us denote the critical value of $\tilde M$ by $\tilde M_*$ and consider the bare quark mass $\tilde m$ as an order parameter, denoting its critical value by $\tilde m_*$. We are interested in calculating the critical exponent $\alpha$ defined by:
 \begin{equation}
| \tilde M-\tilde M_* |\propto|\tilde m-\tilde m_*|^\alpha\ .
\label{critexp}
\end{equation}
We will provide a somewhat heuristic argument that $\alpha=2$ and will confirm this numerically. To begin with let us consider the energy density of the gauge theory $\tilde E$ as a function of the bare quark mass $\tilde m$. Now let us consider a state close to the critical one, characterized by:
\begin{equation}
 \tilde M=\tilde M_*+\delta\tilde M;~~~\tilde m=\tilde m_*+\delta\tilde m;~~~ \tilde E=\tilde E_*+\delta\tilde E;\ .
\end{equation}
 Next we assume that as we approach criticality the variation of $\tilde E$ and $\tilde M$ are proportional to the variation of the energy scale and hence $\delta \tilde E\propto\delta\tilde M$. Therefore we have:
 \begin{equation}
 \frac{\delta \tilde M}{\delta \tilde m}\propto \frac{\delta\tilde E}{\delta \tilde m}\propto \tilde c\ ,
 \label{relation}
 \end{equation}
 where $\tilde c$ is the fermionic condensate. The second relation in (\ref{relation}) was argued in \cite{Kruczenski:2003uq}. In the previous section we argued that the critical exponent of the condensate is one and since the critical embedding has a zero condensate it follows that $\tilde c\propto |\tilde m-\tilde m_{*}|$. Therefore we have:
 \begin{equation} 
  \frac{\delta \tilde M}{\delta \tilde m}\propto \alpha|\tilde m-\tilde m_*|^{\alpha-1}\propto |\tilde m-\tilde m_*|
  \end{equation}
 and hence $\alpha=2$.
 
 Now let us go back to figure \ref{fig:spectrum-spiral}. As we discussed above, for each energy level $n$ the tachyonic spectrum spirals to the critical value $\omega_I^{(n)}$, corresponding to the center of the spiral. If we focus on the $\tilde m=0$ axis, we can see that for each level we have a tower of tachyonic states at a zero bare quark mass, corresponding to the different branches of the spiral. Let us denote by $\tilde M_k^{(n)}$ the imaginary part of the meson spectrum, corresponding to the $k$-th tachyonic state of the $n$-th energy level, at a zero bare quark mass $\tilde m$. As we go deeper into the spiral, $k\to\infty$ and $\tilde M_k^{(n)}\to \tilde M_*^{(n)}$, the data in figure \ref{fig:spectrum-spiral} suggests that $\tilde M_*^{(n)}=\omega_I^{(n)}$. On the other side if the meson spectrum has a critical exponent of two, one can show that for a large $k$:
 \begin{equation}
 \frac{\tilde M_k^{(n)}-\tilde M_*^{(n)}}{\tilde M_{k-1}^{(n)}-\tilde M_*^{(n)}}=q^2\ ,
 \label{geom2}
 \end{equation}
 where $q$ is given by equation (\ref{q}). We can solve for $\tilde M_{*}^{(n)}$:
 \begin{equation}
 \tilde M_{*}^{(n)}=\tilde M_{k-1}+\frac{\tilde M_k^{(n)}-\tilde M_{k-1}^{(n)}}{1-q^2}\ .
 \label{M*}
  \end{equation}
 
Now assuming that for $k=1,2$ the approximate geometrical series defined {\it via} (\ref{geom2}) is already exact we calculate numerically $\tilde M_1^{(n)}, \tilde M_2^{(n)}$ for the $n=0,1,2$ levels and compare the value of $\tilde M_{*}^{(n)}$ obtained by equation (\ref{M*}) to the first three entries in table \ref{tab:1}. The results are presented in table \ref{tab:2}.

\begin{table}[h]
    \caption{}
\begin{center}
\begin{tabular}{|c|c|c|c|c|}
\hline
 $n$&$\tilde M_{1}^{(n)}$&$\tilde M_{2}^{(n)}$&$\tilde M_{*}^{(n)}$&$\omega_I^{(n)}$\\\hline
 0&$2.7530\times10^{-1}$&$2.6460\times10^{-1}$&$2.6447\times10^{-1}$&$2.6448\times10^{-1}$\\\hline
 1&$3.0162\times10^{-2}$&$2.8917\times10^{-2}$&$2.8902\times10^{-2}$&$2.8902\times10^{-2}$\\\hline
 2&$3.2715\times10^{-3}$&$3.1363\times10^{-3}$&$3.1347\times10^{-3}$&$3.1348\times10^{-3}$\\\hline
 
\end{tabular}
\end{center}
\label{tab:2}
\end{table}%

\begin{figure}[h] 
    \centering
    \includegraphics[width=12cm]{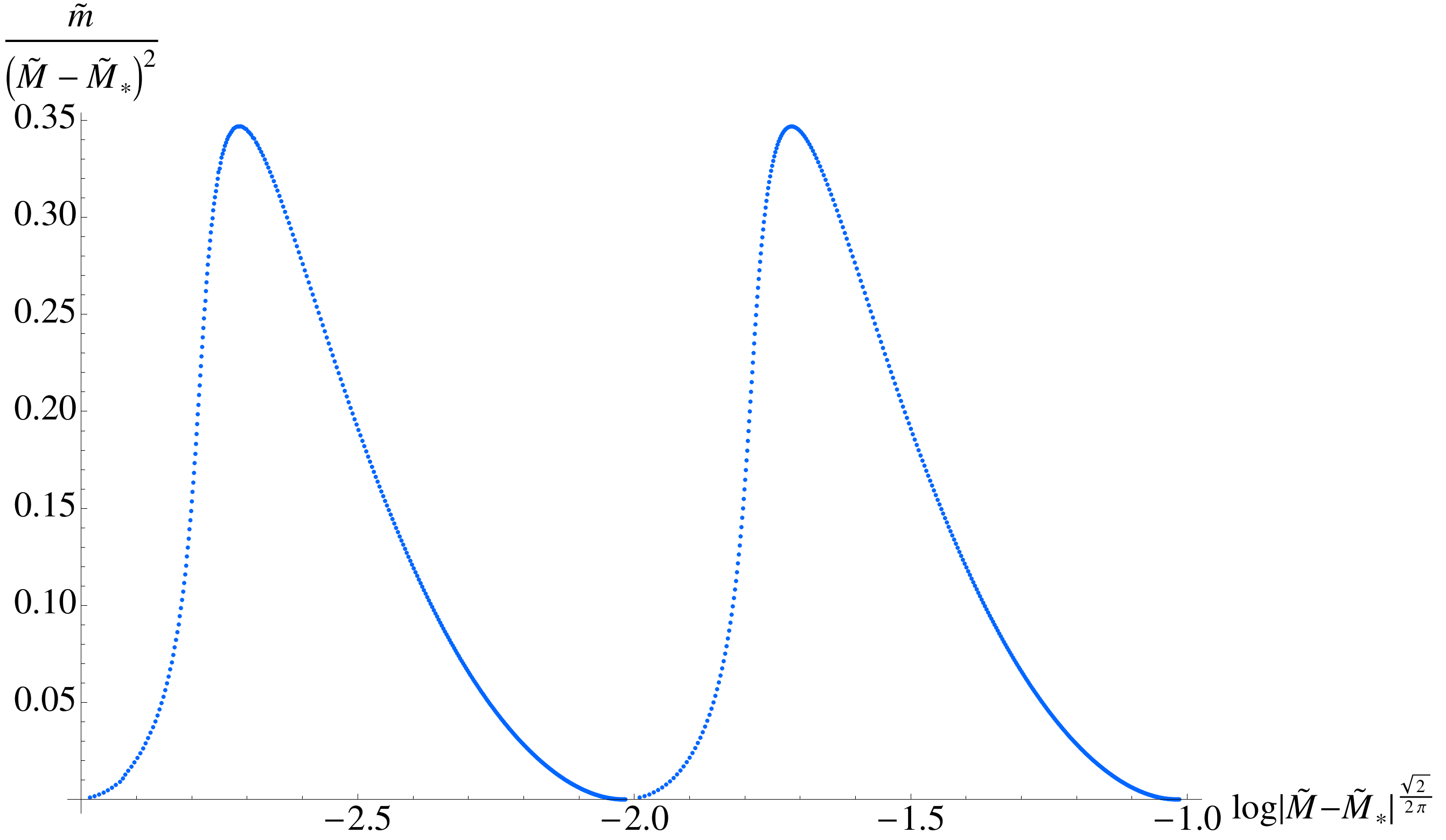} 
    \caption{\small A plot of the bare quark mass meson vs. the meson spectrum, in an appropriate parameterization, determined by the critical exponents of $\tilde m$ and $\tilde M$. The discrete self-similar structure of the spectrum is manifested by the periodicity of the plotted function.}
    \label{fig:messpiral}
 \end{figure}

 One can see that up to four significant digits the critical value of the meson spectrum is given by the imaginary part of the quasi normal modes presented in table \ref{tab:1}. This supports the above argument that the meson spectrum has a critical exponent of two. Another way to justify this, is to generate a plot of the meson spectrum similar to the one presented in figure \ref{fig:mspiral} for the bare quark mass $\tilde m$ and the fermionic spectrum $\tilde c$. Notice that $\tilde M$ approaches criticality from above, while the parameter $\tilde m$ oscillates around the critical value $\tilde m_*=0$. This suggests to use $\tilde M$ as an order parameter and to generate a plot of $\tilde m/(\tilde M-\tilde M_*)^2$ vs. $\sqrt{2}\log{|\tilde M-\tilde M_*|}/{2\pi}$. Note that according to equation (\ref{geom2}) the plot should represent periodic function of an unit period. The resulting plot for the $n=0$ level, using $\tilde M_*^{(0)}$ from table \ref{tab:2} as a critical value, is presented in figure \ref{fig:messpiral}. 
   
\subsubsection{The stable branch of the spiral}   
In this subsection we consider the spectrum corresponding to the states far from the origin of the $(\tilde m, -\tilde c)$, which is the outermost branch of the spiral ending at point $H_0$ from figure \ref{fig:spiral-revisited}. The fluctuations of the D7-brane corresponding to the massless scalar $\phi$ were studied in \cite{Filev:2007gb} and some features consistent with the spontaneous chiral symmetry breaking, such as a characteristic $\sqrt{m}$ behavior \cite{Gell-Mann:1968rz} were reported.  

 Here we complement the analysis by presenting the results for the fluctuations along the $\tilde L$ coordinate. Since this is the massive field in the spontaneous chiral symmetry breaking scenario, we  expect a $\sqrt{const+\tilde m}$ behavior of the meson spectrum for small values of $\tilde m$. Note that such a behavior simply means that the spectrum of the $\tilde L$ fluctuations has a mass gap at zero bare quark mass and that the slope of the spectrum vs. the bare quark mass function is finite. It is satisfying that our results are in accord with this expectations. 
  
\begin{figure}[h] 
   \centering
   \includegraphics[width=12cm]{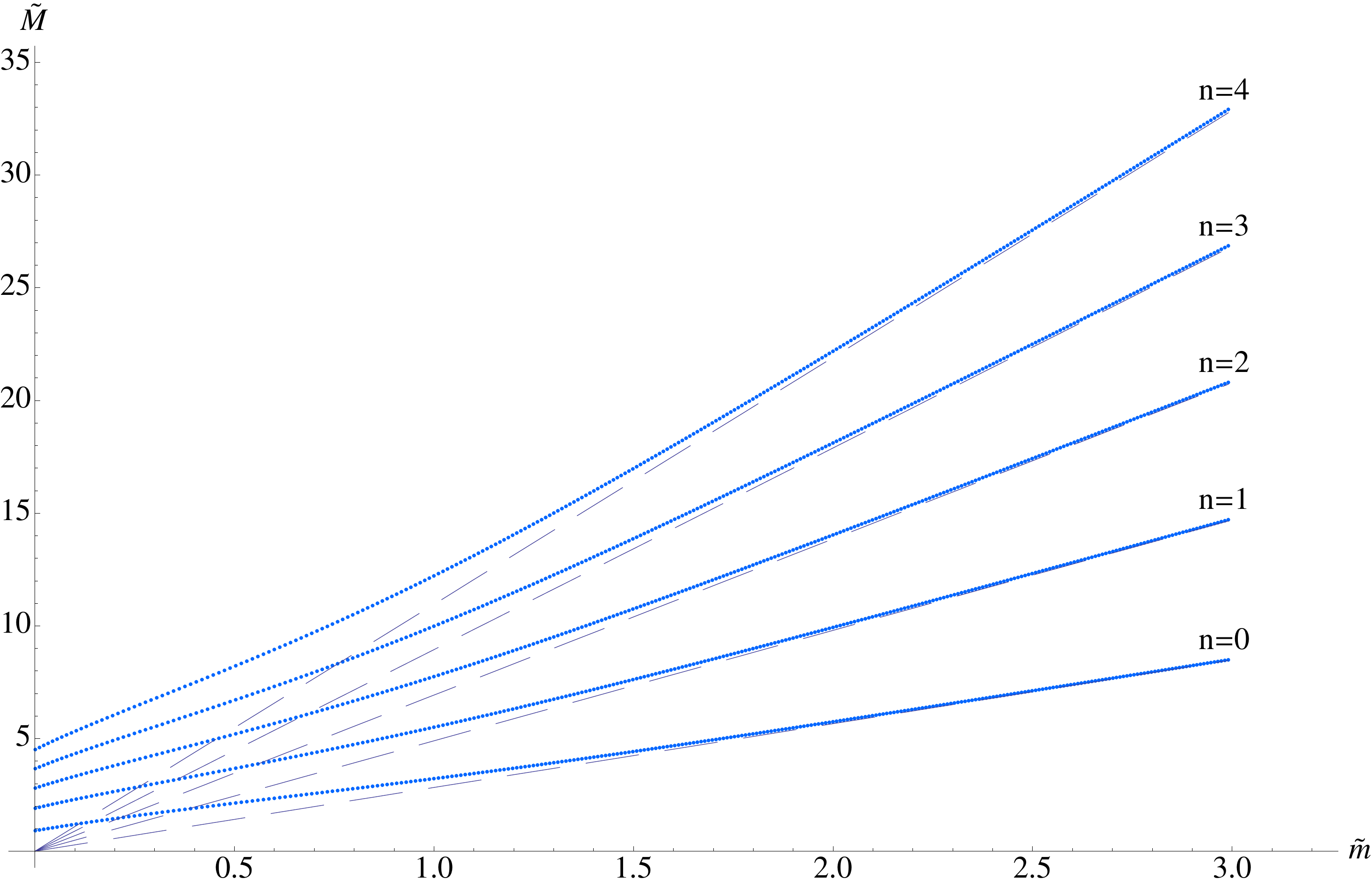} 
   \caption{\small A plot of the meson spectrum corresponding to the stable branch of the spiral. The black dashed lines correspond to equation (\ref{dmlAdS}), one can see that for large $\tilde m$ the meson spectrum asymptotes to the result for pure $AdS_5\times S^5$ space. One can also see that at zero bare quark mass $\tilde m$ there is a mass gap in the spectrum. 
   }
   \label{fig:final}
\end{figure}

 To obtain the spectrum, we solve numerically equation (\ref{fluctPsi}) imposing Neumann boundary conditions at $\tilde\rho=0$. A plot of the first five energy levels is presented in figure \ref{fig:final}. As one can see at large $\tilde m$ the spectrum approximates that of the pure ${\cal N}=2$ Flavored Yang Mills theory studied in \cite{Kruczenski:2003be}, where the dependence of the meson spectrum on the bare quark mass was obtained in a closed form:
 \begin{equation}
M_0=\frac{2m}{R^2}\sqrt{(n+l+1)(n+l+2)}\ .
\label{spectrAds}
\end{equation}
Here $l$ is the quantum number corresponding to the angular modes along the internal $S^3$ sphere wrapped by the D7 brane and is zero in our case. After introducing the dimensionless variables defined in (\ref{dimensionless}), equation (\ref{spectrAds}) boils down to:
\begin{equation}
\tilde M_0= 2\sqrt{(n+1)(n+2)}\tilde m\ .
\label{dmlAdS}
\end{equation}
The black dashed lines in figure \ref{fig:final} represent equation (\ref{dmlAdS}). The fact that the meson spectrum asymptotes to the one described by (\ref{dmlAdS}) justifies the use of the quantum number $n$ to classify the meson spectrum. One can also see that as expected the spectrum at zero bare quark mass has a mass gap.

\section{Conclusion}

In this paper we performed a detailed analysis of the spiral structure at the origin of the condensate vs. bare quark mass diagram. We revealed the discrete self-similar behavior of the theory near criticality and calculated the corresponding critical exponents for the bare quark mass, the fermionic condensate and the meson spectrum.

 Our study of the meson spectrum confirmed the expectations based on thermodynamic considerations that the lowest positive $\tilde m$ branch of the spiral corresponds to a stable phase of the theory and that  the inner branches are real instabilities characterized by a tachyonic ground state and cannot be reached by a supercooling. The lowest negative $\tilde m$ branch of the spiral is tachyon free and thus could be metastable.
 
The supercooling mentioned above could be attempted by considering the finite temperature background, namely the AdS Black hole geometry, in the presence of an external magnetic field. We could prepare the system in the phase corresponding to the trivial $\tilde L\equiv 0$ embedding and then take the $T\to0$ limit. If some of the inner branches of the spiral were metastable the theory could end up in the corresponding phase. The study of the finite temperature case is of a particular interest. Due to the additional scale introduced by the temperature, the theory has two dimensionless parameters and is described by a two dimensional phase diagram. The effect of the temperature is to restore the chiral symmetry and is competing with that of the external magnetic field. On the other side the magnetic field affects the melting of the mesons \cite{Albash:2007bk}.

\section{Acknowledgments}
V. Filev would like to thank: T. Albash, C. V. Johnson, A. Kundu and R. Rashkov for useful comments and discussions. This work was supported in part by the US Department of Energy. 
\newpage
\appendix \section{Calculating the condensate of the theory.} \label{appendix:A}
Let us consider the on-shell action:
\begin{equation}
S= -2\pi^2\tau_7N_f\int\limits_0^{\rho_{max}}d\rho\rho^3\sqrt{1+\frac{R^4H^2}{\rho^2+L(\rho)^2}}\sqrt{1+L'(\rho)^2}
\label{Saction}
\end{equation}
it diverges as $\rho_{max}\to\infty$. To rectify this we regularize the action by subtracting the action for the $L\equiv 0$ embedding:
\begin{eqnarray}
&&\frac{S_{sub}}{2\pi^2\tau_7N_f}=-\int\limits_0^{\rho_{max}}d\rho\rho^3\sqrt{1+\frac{R^4H^2}{\rho^4}}\frac{1}{4}\rho_{max}^2\sqrt{\rho_{max}^4+R^4H^2}+\\
&&\frac{1}{4}R^4H^2\ln{\left(\frac{\rho_{max}^2+\sqrt{R^4H^2+\rho_{max}^4}}{R^2H}\right)}\nonumber
\end{eqnarray}
 This results to the following regularized action:
\begin{equation}
S_{reg}[L,\rho_{max}]=S-S_{sub}
\label{Sreg}
\end{equation}
On the other-side the hamiltonian density of the theory can be written as \cite{Kruczenski:2003uq}:
\begin{equation}
{\cal H}=\int d^2\theta m_q{\tilde Q}Q+{\cal H}_{0}\ ,
 \end{equation}
 where ${\cal H}_0$ is the mass independent part of the hamiltonian and $Q,\tilde Q$ are the two chiral fields of the hypermultiplet in ${\cal N}=1$ notations. Now by making the identification:
 \begin{equation}
 \langle{\cal H}\rangle=-\lim_{\rho_{max}\to\infty}S_{reg}
 \end{equation}
and using equations (\ref{Saction})-(\ref{Sreg}) and the asymptotic of $L(\rho)$ as $\rho_{max}\to\infty$:
\begin{equation}
L(\rho)=m+\frac{c}{\rho_{max}^2}+\dots\ ,
\end{equation}
we can obtain:
 \begin{eqnarray}
\langle\frac{\delta{\cal H}}{\delta m_q}\rangle&=&\langle\bar\psi\psi\rangle=-2\pi\alpha'\lim_{\rho_{max}\to\infty}\frac{\delta S_{reg}}{\delta L}=\\
&=&\lim_{\rho_{max}\to\infty}4\pi^3\alpha'\tau_7 N_f\rho_{max}^3L'(\rho_{max})=-8\pi^3\alpha'\tau_7N_f c\ .\nonumber
 \end{eqnarray}
 And hence we finally get:
\begin{equation}
 \langle\bar\psi\psi\rangle=-\frac{N_f}{(2\pi\alpha')^3g_{YM}^2} c
\end{equation}


\newpage

\begin{thebibliography}{99}
  
 
 
 
 
\bibitem{Aharony:1999ti}
  O.~Aharony, S.~S.~Gubser, J.~M.~Maldacena, H.~Ooguri and Y.~Oz,
  Phys.\ Rept.\  {\bf 323}, 183 (2000)
  [arXiv:hep-th/9905111].

\bibitem{Karch:2002sh}
  A.~Karch and E.~Katz,
  JHEP {\bf 0206}, 043 (2002)
  [arXiv:hep-th/0205236].

\bibitem{Kruczenski:2003be}
M.~Kruczenski, D.~Mateos, R.~C. Myers, and D.~J. Winters, ``Meson spectroscopy
  in AdS/CFT with flavour,'' {\em JHEP} {\bf 07} (2003) 049,


\bibitem{Babington:2003vm}
  J.~Babington, J.~Erdmenger, N.~J.~Evans, Z.~Guralnik and I.~Kirsch,
   ``Chiral symmetry breaking and pions in non-supersymmetric gauge /  gravity
  Phys.\ Rev.\ D {\bf 69}, 066007 (2004)
  [arXiv:hep-th/0306018].

\bibitem{Frolov:2006tc}
  V.~P.~Frolov,
  Phys.\ Rev.\  D {\bf 74}, 044006 (2006)
  [arXiv:gr-qc/0604114].

\bibitem{Kruczenski:2003uq}
  M.~Kruczenski, D.~Mateos, R.~C.~Myers and D.~J.~Winters,
  JHEP {\bf 0405}, 041 (2004)
  [arXiv:hep-th/0311270].


\bibitem{Albash:2006ew}
  T.~Albash, V.~Filev, C.~V.~Johnson and A.~Kundu,
  [arXiv:hep-th/0605088].

\bibitem{Albash:2006bs}
  T.~Albash, V.~Filev, C.~V.~Johnson and A.~Kundu,
   ``Global currents, phase transitions, and chiral symmetry breaking in large
  [arXiv:hep-th/0605175].

\bibitem{Filev:2007gb}
  V.~G.~Filev, C.~V.~Johnson, R.~C.~Rashkov and K.~S.~Viswanathan,
  arXiv:hep-th/0701001.

\bibitem{Arean:2005ar}
  D.~Arean, A.~Paredes and A.~V.~Ramallo,
  JHEP {\bf 0508}, 017 (2005)
  [arXiv:hep-th/0505181].


\bibitem{Strassler}
  S.~Hong, S.~Yoon and M.~J.~Strassler,
  JHEP {\bf 0404}, 046 (2004)
  [arXiv:hep-th/0312071].

\bibitem{Evans:2004ia}
  N.~J.~Evans and J.~P.~Shock,
  Phys.\ Rev.\ D {\bf 70}, 046002 (2004)
  [arXiv:hep-th/0403279].

\bibitem{Ghoroku:2004sp}
  K.~Ghoroku and M.~Yahiro,
  Phys.\ Lett.\ B {\bf 604}, 235 (2004)
  [arXiv:hep-th/0408040].

\bibitem{Erdmenger:2004dk}
  J.~Erdmenger and I.~Kirsch,
  JHEP {\bf 0412}, 025 (2004)
  [arXiv:hep-th/0408113].

\bibitem{Kruczenski:2004me}
  M.~Kruczenski, L.~A.~P.~Zayas, J.~Sonnenschein and D.~Vaman,
  JHEP {\bf 0506}, 046 (2005)
  [arXiv:hep-th/0410035].

\bibitem{Kuperstein:2004hy}
  S.~Kuperstein,
  JHEP {\bf 0503}, 014 (2005)
  [arXiv:hep-th/0411097].

\bibitem{Sakai:2004cn}
  T.~Sakai and S.~Sugimoto,
  Prog.\ Theor.\ Phys.\  {\bf 113}, 843 (2005)
  [arXiv:hep-th/0412141];
  T.~Sakai and S.~Sugimoto,
  Prog.\ Theor.\ Phys.\  {\bf 114}, 1083 (2006)
  [arXiv:hep-th/0507073].

\bibitem{Ghoroku:2005tf}
  K.~Ghoroku, T.~Sakaguchi, N.~Uekusa and M.~Yahiro,
  Phys.\ Rev.\ D {\bf 71}, 106002 (2005)
  [arXiv:hep-th/0502088].

\bibitem{Vaman}
  I.~Kirsch and D.~Vaman,
  Phys.\ Rev.\ D {\bf 72}, 026007 (2005)
  [arXiv:hep-th/0505164].

\bibitem{Brevik:2005fs}
  I.~Brevik, K.~Ghoroku and A.~Nakamura,
  Int.\ J.\ Mod.\ Phys.\ D {\bf 15}, 57 (2006)
  [arXiv:hep-th/0505057].

\bibitem{Levi:2005hh}
  T.~S.~Levi and P.~Ouyang,
  arXiv:hep-th/0506021.

\bibitem{Peeters:2005fq}
  K.~Peeters, J.~Sonnenschein and M.~Zamaklar,
  JHEP {\bf 0602}, 009 (2006)
  [arXiv:hep-th/0511044].

\bibitem{Cotrone:2005fr}
  A.~L.~Cotrone, L.~Martucci and W.~Troost,
  Phys.\ Rev.\ Lett.\  {\bf 96}, 141601 (2006)
  [arXiv:hep-th/0511045].

\bibitem{Bigazzi:2006jt}
  F.~Bigazzi and A.~L.~Cotrone,
  arXiv:hep-th/0606059.


\bibitem{Ghoroku:2005kg}
  K.~Ghoroku and M.~Yahiro,
  Phys.\ Rev.\ D {\bf 73}, 125010 (2006)
  [arXiv:hep-ph/0512289].

\bibitem{Arean:2006pk}
  D.~Arean and A.~V.~Ramallo,
  JHEP {\bf 0604}, 037 (2006)
  [arXiv:hep-th/0602174].

\bibitem{Antonyan:2006vw}
  E.~Antonyan, J.~A.~Harvey, S.~Jensen and D.~Kutasov,
  arXiv:hep-th/0604017.

\bibitem{Mateos:2006nu}
  D.~Mateos, R.~C.~Myers and R.~M.~Thomson,
  arXiv:hep-th/0605046.


\bibitem{Aharony:2006da}
  O.~Aharony, J.~Sonnenschein and S.~Yankielowicz,
  arXiv:hep-th/0604161.


\bibitem{Erdmenger:2006bg}
  J.~Erdmenger, N.~Evans and J.~Grosse,
  arXiv:hep-th/0605241.

\bibitem{Kirsch:2006he}
  I.~Kirsch,
  arXiv:hep-th/0607205.

\bibitem{Herzog:2006se}
  C.~P.~Herzog,
  arXiv:hep-th/0605191.

\bibitem{Herzog:2006gh}
  C.~P.~Herzog, A.~Karch, P.~Kovtun, C.~Kozcaz and L.~G.~Yaffe,
   ``Energy loss of a heavy quark moving through N = 4 supersymmetric Yang-Mills
  JHEP {\bf 0607}, 013 (2006)
  [arXiv:hep-th/0605158].

\bibitem{Gao:2006up}
  Y.~h.~Gao, W.~s.~Xu and D.~f.~Zeng,
  JHEP {\bf 0608}, 018 (2006)
  [arXiv:hep-th/0605138].

\bibitem{Karch:2006bv}
  A.~Karch and A.~O'Bannon,
  arXiv:hep-th/0605120.

\bibitem{Horigome:2006xu}
  N.~Horigome and Y.~Tanii,
  arXiv:hep-th/0608198.

\bibitem{Matsuo:2006ws}
  T.~Matsuo, D.~Tomino and W.~Y.~Wen,
  arXiv:hep-th/0607178.

\bibitem{Apreda:2006bu}
  R.~Apreda, J.~Erdmenger, D.~Lust and C.~Sieg,
  arXiv:hep-th/0610276.

\bibitem{Nakamura:2006xk}
  S.~Nakamura, Y.~Seo, S.~J.~Sin and K.~P.~Yogendran,
  arXiv:hep-th/0611021.

\bibitem{Kobayashi:2006sb}
  S.~Kobayashi, D.~Mateos, S.~Matsuura, R.~C.~Myers and R.~M.~Thomson,
  arXiv:hep-th/0611099.

\bibitem{Evans:2007we}
  N.~Evans,
  arXiv:hep-ph/0701218.
  
\bibitem{Hoyos:2006gb}
  C.~Hoyos, K.~Landsteiner and S.~Montero,
  JHEP {\bf 0704}, 031 (2007)
  [arXiv:hep-th/0612169].

\bibitem{Buchel:2007vy}
  A.~Buchel, S.~Deakin, P.~Kerner and J.~T.~Liu,
  arXiv:hep-th/0701142.

\bibitem{Mateos:2007vn}
  D.~Mateos, R.~C.~Myers and R.~M.~Thomson,
  arXiv:hep-th/0701132.

\bibitem{Erdmenger:2007ap}
  J.~Erdmenger, M.~Kaminski and F.~Rust,
  arXiv:0704.1290 [hep-th].
  
\bibitem{Erdmenger:2007bn}
  J.~Erdmenger, R.~Meyer and J.~P.~Shock,
  JHEP {\bf 0712}, 091 (2007)
  [arXiv:0709.1551 [hep-th]].

\bibitem{Albash:2007bk}
  T.~Albash, V.~G.~Filev, C.~V.~Johnson and A.~Kundu,
  arXiv:0709.1547 [hep-th].

\bibitem{Karch:2007pd}
  A.~Karch and A.~O'Bannon,
  arXiv:0705.3870 [hep-th].

\bibitem{Myers:2007we}
  R.~C.~Myers, A.~O.~Starinets and R.~M.~Thomson,
  arXiv:0706.0162 [hep-th].

\bibitem{Albash:2007bk}
  T.~Albash, V.~G.~Filev, C.~V.~Johnson and A.~Kundu,
  arXiv:0709.1547 [hep-th].



\bibitem{Gell-Mann:1968rz}
  M.~Gell-Mann, R.~J.~Oakes and B.~Renner,
  Phys.\ Rev.\  {\bf 175} (1968) 2195.

\bibitem{Johnson:2003gi}
  C.~V.~Johnson,
  ``D-Branes,'' Cambridge University Press, 2003.



\end{thebibliography}
\end{document}